\documentclass[12pt,preprint]{aastex}

\begin{document}

\newcommand{\nhires}{42}
\newcommand{\nesi}{39}
\newcommand{\ntot}{65}
\newcommand{\ndla}{86}
\newcommand{\nmtl}{153}
\newcommand{\newdla}{10}
\newcommand{\kms}{km~s$^{-1}$ }
\newcommand{\cm}[1]{\, {\rm cm^{#1}}}
\newcommand{\mkms}{{\rm \; km\;s^{-1}}}
\newcommand{\delv}{\Delta v}
\newcommand{\ohi}{$\Omega_g$}
\newcommand{\lya}{Ly$\alpha$}
\newcommand{\nv}{N\,V}
\newcommand{\ovi}{O\,VI}
\newcommand{\N}[1]{{N({\rm #1})}}
\newcommand{\sci}[1]{{\rm \; \times \; 10^{#1}}}
\newcommand{\mnhi}{N_{\rm HI}}
\newcommand{\mnciv}{N_{\rm CIV}}
\newcommand{\nhi}{$N_{\rm HI}$}
\def\fnhi{$f_{\rm{HI}} (\mnhi)$}
\def\mfnhi{f_{\rm{HI}} (\mnhi)}
\def\ltk{\left [ \,}
\def\ltp{\left ( \,}
\def\ltb{\left \{ \,}
\def\rtk{\, \right  ] }
\def\rtp{\, \right  ) }
\def\rtb{\, \right \} }
\def\nhi{$N_{\rm HI}$}
\def\lnhi{$\log N_{HI}$}
\def\omt{$\Omega_m^{Total}$}
\def\momt{\Omega_m^{Total}}

\title{THE UCSD/KECK DAMPED LY$\alpha$ ABUNDANCE DATABASE: A Decade of 
High Resolution Spectroscopy}

\def\ucsc{2}
\def\ucsd{3}
\def\und{4}
\def\yale{5}
\def\mit{6}
\def\uci{7}
\def\hamb{8}

\author{Jason X. Prochaska\altaffilmark{1,\ucsc},
Arthur M. Wolfe\altaffilmark{1,\ucsd}, 
J. Christopher Howk\altaffilmark{1,\und},
Eric Gawiser\altaffilmark{1,\yale},
Scott M. Burles\altaffilmark{1,\mit},
Jeff Cooke\altaffilmark{1,\uci}
}

\altaffiltext{1}{Visiting Astronomer, W.M. Keck Telescope.
The Keck Observatory is a joint facility of the University
of California, California Institute of Technology, and NASA.}
\altaffiltext{\ucsc}{Department of Astronomy and Astrophysics, 
UCO/Lick Observatory;
University of California, 1156 High Street, 
Santa Cruz, CA 95064; xavier@ucolick.org}
\altaffiltext{\ucsd}{Department of Physics, and Center for Astrophysics and 
Space Sciences, University of California, San Diego, C--0424, La Jolla, 
CA 92093-0424}
\altaffiltext{\und}{Department of Physics, University of Notre Dame, Notre Dame, IN 46556}
\altaffiltext{\yale}{NSF Astronomy and Astrophysics Postdoctoral Fellow, 
Yale Astronomy Department and Yale Center for Astronomy and Astrophysics, 
PO Box 208101, New Haven, CT 06520}
\altaffiltext{\mit}{MIT Kavli Institute for Astrophysics and Space Research,Massachusetts Institute of Technology, 77 Massachusetts Avenue, Cambridge MA 02139}
\altaffiltext{\uci}{Department of Physics and Astronomy and Center 
for Cosmology, University
of California, Irvine, 4129 Frederick Reines Hall, 
Irvine, CA 92697-4575; cooke@uci.edu}

\begin{abstract}
We publish the Keck/HIRES and Keck/ESI spectra that we have obtained
during the first 10 years of Keck observatory operations.
Our full sample includes \nhires\ HIRES spectra and \nesi\ ESI spectra along
\ntot\ unique sightlines providing abundance measurements on 
$\approx 85$ DLA systems.  
The normalized data can be downloaded from the journal or 
from our supporting website:
http://www.ucolick.org/$\sim$xavier/DLA/.
The database includes all of the sightlines that have been
included in our papers on the chemical abundances, kinematics,
and metallicities of the damped \lya\ systems.
This data has also been used to argue for variations in the
fine-structure constant.
We present new chemical abundance measurements for \newdla\
damped \lya\ systems and a summary table of high-resolution
metallicity measurements (including values from the literature)
for \nmtl\ damped \lya\ systems at $z>1.6$.
We caution, however, that this metallicity sample (and all previous
ones) is biased to higher \nhi\ values than a random sample.
\end{abstract}


\section{Introduction}

Since the discovery of the damped \lya\ (DLA) systems \citep{wolfe86},
researchers have appreciated that high resolution spectra gives
special insight into the interstellar medium (ISM) of high $z$ galaxies
in a fashion analogous with the local universe \citep[e.g.][]{ss96,welty99}.
For 8 years following this initial work, however, 
astronomers had access to only 4m-class
telescopes and echelle observations were limited to the brightest
quasars \citep[e.g.][]{dodorico91,tls96}.
The commissioning of the High Resolution Echelle Spectrometer 
\citep[HIRES;][]{vogt94} on the Keck~I 10m telescope heralded a new era
in the study of high $z$ galaxies \citep{wolfe94,wgp05}.
The years that followed witnessed the commissioning of additional
echelle spectrometers on 10m-class telescopes -- VLT/UVES \citep{uves}, 
Magellan/MIKE 
\citep{bernstein03}, Subaru/HDS \citep{hds} -- and
the commissioning of an echellette spectrometer on the Keck~II 10m telescope
\citep[ESI][]{sheinis02}).
These technological advances have led to the research of many topics, including
the kinematic characteristics \citep{pw97}, metallicities \citep{pw00,ledoux06},
depletion patterns \citep{lu96,pettini00,dz06}, nucleosynthetic abundances
\citep{lu96,molaro00,mirka01}, molecular fraction \citep{petit00,ledoux03},
and interstellar physics \citep{wpg03,howk05} for a 
sample of high redshift galaxies which are representative of the
full galaxy distribution. 

In this paper, we publish the spectra obtained by our group using the
HIRES and ESI spectrometers during the first decade of Keck operations.
Our full sample includes \nhires\ HIRES spectra and \nesi\ ESI spectra along
\ntot\ unique sightlines providing abundance measurements on 
\ndla\ DLA systems.  Previous papers have analyzed this data
for many scientific purposes, and we refer the reader to those papers
for further discussions:  
(1) gas kinematics \citep{pw97,pw98,wp98,wp00a,pw01};
(2) metallicity measurements \citep{pw00,pgw+03}; 
(3) chemical abundances \citep{pw99,pw02,phw03};
(4) nitrogen enrichment \citep{pho+02};
(5) photoionization \citep{pro_ion02};
(6) star formation rates \citep{wpg03,wgp03,w04};
(7) interstellar physics \citep{howk05};
(8) variations of the fine-structure constant \citep{webb01,murphy01b};
and 
(9) a survey of super-Lyman Limit systems \citep{omeara06}.

Previously, the data has been proprietary and, unfortunately, current
plans for the Keck Observatory Archive 
(KOA\footnote{http://msc.caltech.edu/archives/koa/})
do not include this dataset.\footnote{Note that all HIRES data
obtained by our group with the CCD mosaic (i.e. post September 2004)
will be archived by KOA.}
And, while we have mined these spectra extensively, we believe there are
still new discoveries to be made.
The ESI data, in particular, continue to offer new avenues of research,
e.g., a molecular hydrogen survey (Milutinovich et al, in prep),
studies of the \lya\ forest (Abazajian et al., in prep), and 
an extensive survey for [C II] cooling rates (Wolfe et al., in prep).
This paper and its associated on-line resources will make our full
set of DLA observations freely available to the general astronomical
community.
The paper is summarized as follows. $\S$~\ref{sec:obs} 
briefly describes the observations
and data reduction procedures.  $\S$~\ref{sec:abund} presents 
Figures and Tables of previously unpublished (by our group)
DLA transitions and abundance measurements.
Finally, $\S$~\ref{sec:summ} provides a brief summary.

\section{SUMMARY OF OBSERVATIONS AND DATA REDUCTION}
\label{sec:obs}

Most of the spectra presented here have been previously
summarized in three papers \citep{pw99,pro01,p03_esi}, but for
completeness we provide a journal of all the observations in 
Tables~\ref{tab:hires} and \ref{tab:esi}, for HIRES and ESI
respectively.
For nearly all observations, the HIRES spectra were acquired using
either a $0.8''$ or $1.1''$ wide decker 
(FWHM~$\approx 6$ and 8 \kms, respectively)
and the ESI observations were carried out with the 0.5$''$ or
$0.75''$ slit (FWHM~$\approx 33$ and 44 \kms, respectively).
All of the HIRES spectra were acquired with the original
Tektronix 2048$\times$2048 CCD.
As such, we were reluctant to observe at wavelengths bluer than
$\lambda \approx 4000$\AA\ or redward of $\lambda \approx 8500$\AA.
In general we strove to achieve a final signal-to-noise (S/N) ratio
of $>15$ per pixel (2 \kms pix$^{-1}$ for HIRES, 11 \kms pix$^{-1}$ for ESI).
ESI has a fixed format which covers the spectral region
$\lambda = 4000$ to $10,000$\AA.

With only two exceptions (Q1331+17, PHL957), all
of the HIRES data were reduced with various versions of the 
MAKEE package,\footnote{http://spider.ipac.caltech.edu/staff/tab/makee/} 
kindly developed and distributed by T. Barlow.  
This data reduction pipeline bias subtracts and flattens the 2D images
using standard techniques.  It traces the object 
with the object itself, a standard star, or a pinhole spectrum
of the quartz lamp.  
The data is sky-subtracted and optimally extracted.
All but the original produced a 1D
wavelength solution for each echelle order.  The 1D spectra were coadded
with various in-house algorithms that weighted by S/N ratio and
rejected spurious pixels.  In most cases, the data were first
normalized using the {\it xplot} routine within MAKEE.
This procedure was relatively straightforward because little of
this HIRES data includes coverage the \lya\ forest.
The complete set of normalized, 1D spectra and error arrays are available
with the electronic version of this paper and also at this online archive: 
http://www.ucolick.org/$\sim$xavier/DLA/.
Note that the gaps in the spectra are echelle order gaps and occur
redward of $\approx 5200$\AA.

All of the ESI observations were reduced with the ESIRedux package 
developed in IDL by J.X. Prochaska \citep{p03_esi},
which is publically 
available.\footnote{http://www2.keck.hawaii.edu/inst/esi/ESIRedux/index.html}
The majority of data were extracted with a simple boxcar aperture.
Finally, the spectra are fluxed (erg s$^{-1}$ cm$^{-2}$ \AA$^{-1}$)
and fluxed with an archived sensitivity function.  
The flux is reasonably accurate ($\approx 20\%$) in a relative sense
but we have made no corrections for slit loss, airmass, or Galactic
reddening.   A continuum fit was
derived for the data redward of the quasar \lya\ emission feature
using the {\it x\_continuum} routine within the XIDL 
package.\footnote{http://www.ucolick.org/$\sim$xavier/IDL/index.html}  
Both the fluxed and normalized spectra are available online.
The spectra are continuous with the exception of the data near 
4500\AA\ where a chip blemish removes approx 50\AA\ of data.

\section{NEW ABUNDANCE MEASUREMENTS}
\label{sec:abund}

\begin{figure}
\epsscale{0.85}
\plotone{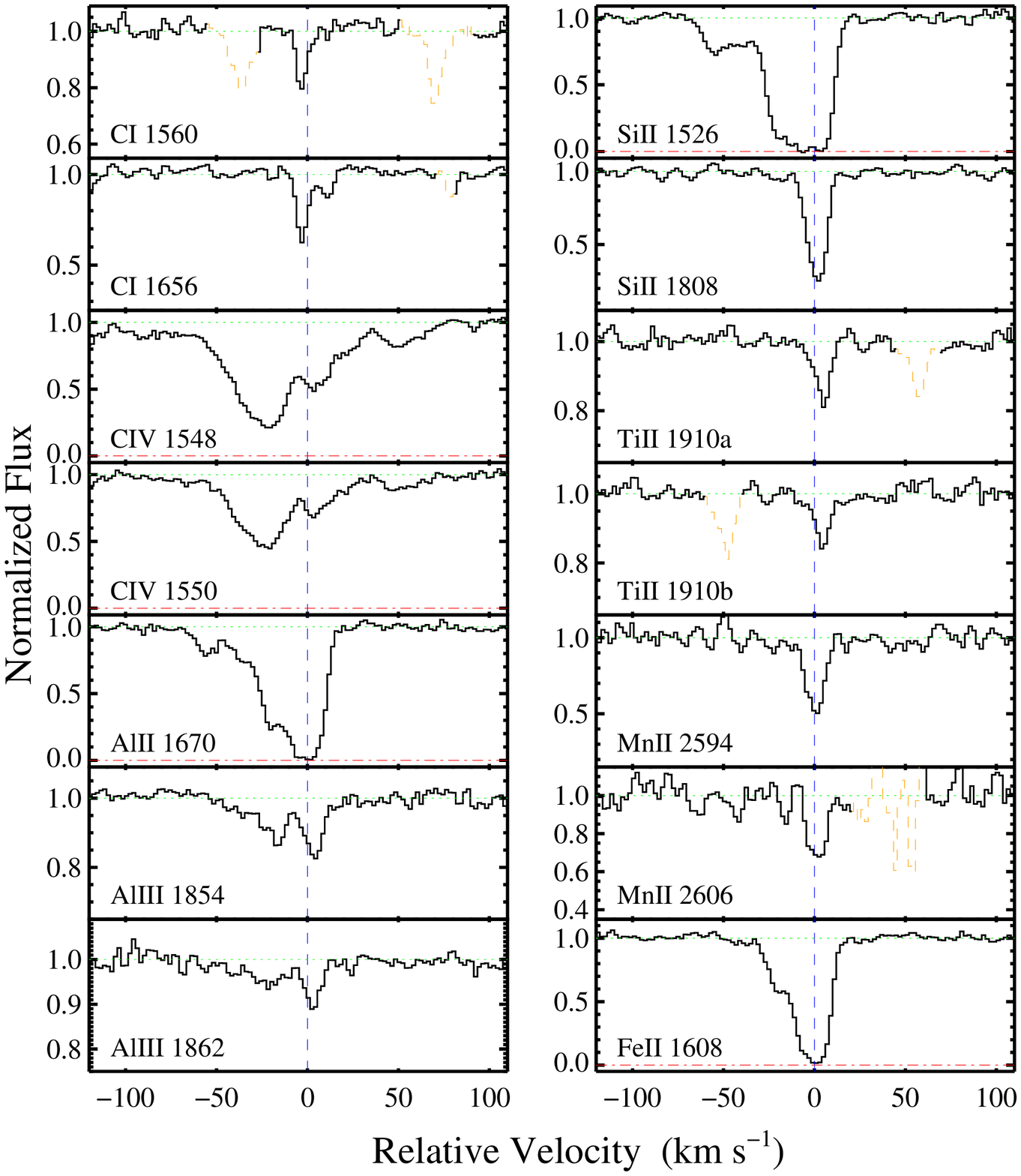}
\caption{HIRES velocity profiles of the transitions identified
with the damped \lya\ systems at $z=2.0668$ toward FJ0812+32.
}
\label{fig:fj0812_z206}
\end{figure}

\begin{figure}
\epsscale{0.85}
\plotone{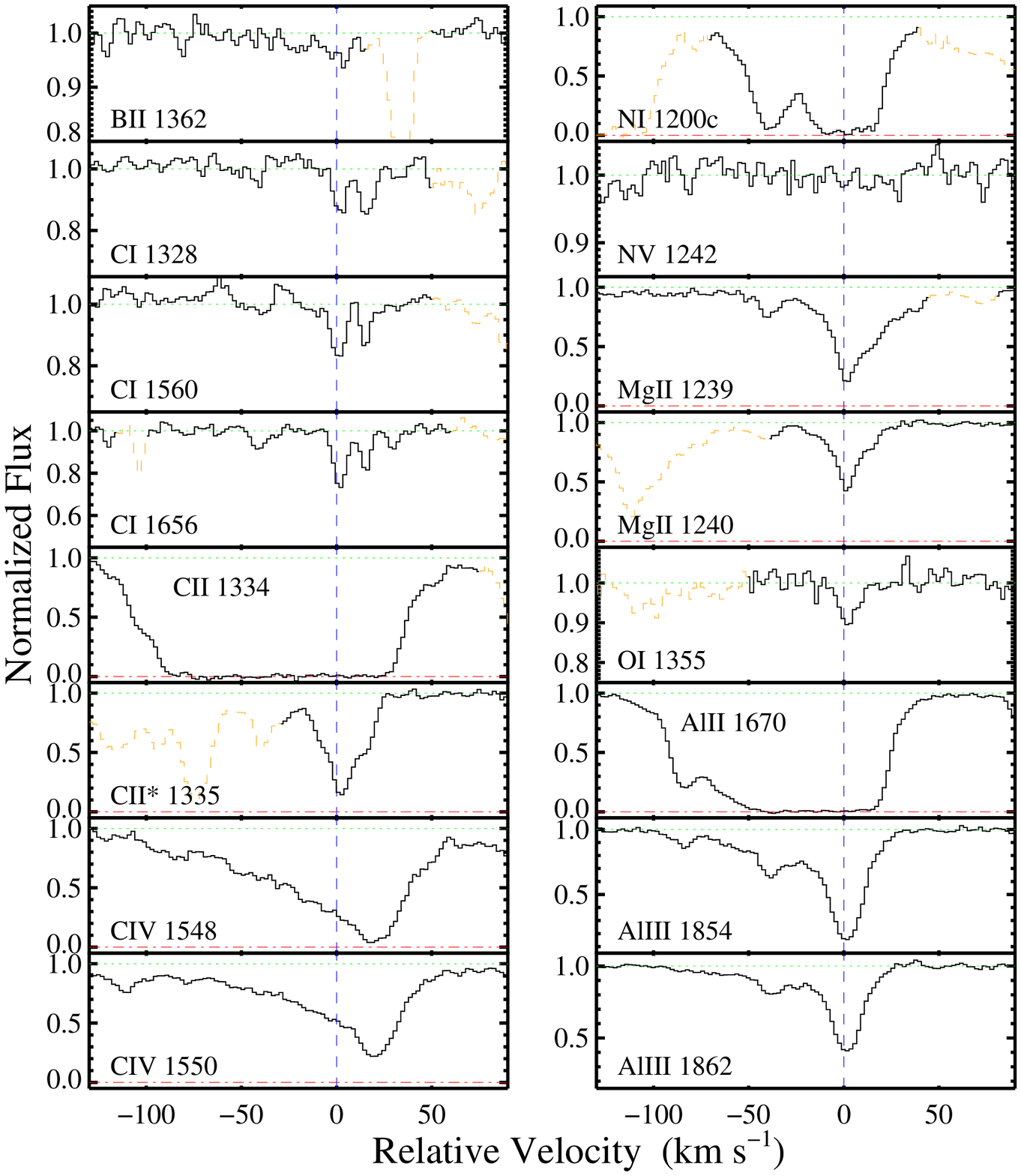}
\caption{HIRES velocity profiles of the transitions identified
with the damped \lya\ systems at $z=2.6263$ toward FJ0812+32.
}
\label{fig:fj0812_z262}
\end{figure}

\begin{figure}
\epsscale{0.85}
\plotone{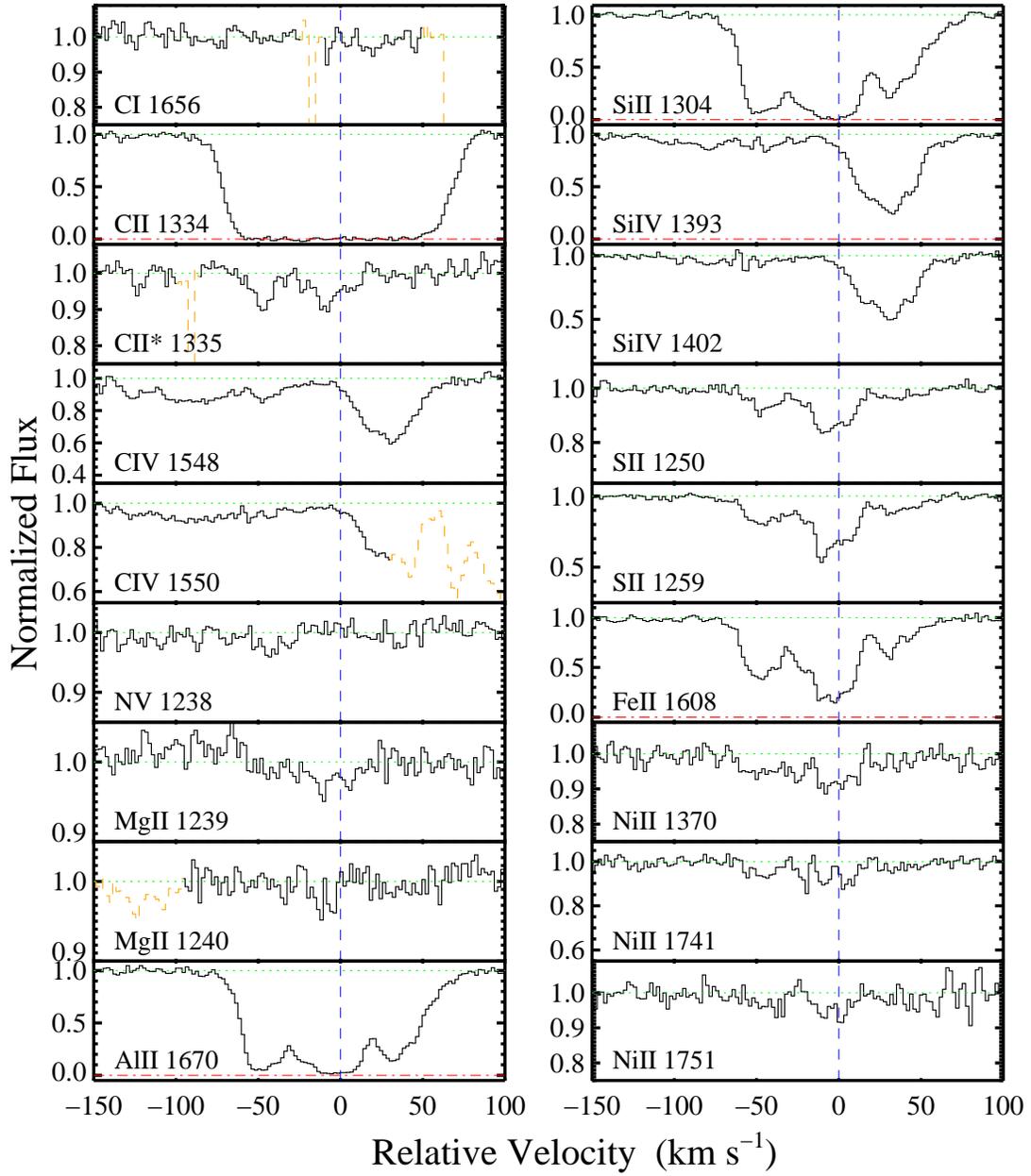}
\caption{HIRES velocity profiles of the transitions identified
with the damped \lya\ systems at $z=3.2458$ toward J0900+42.
}
\label{fig:j0900_z324}
\end{figure}

\begin{figure}
\epsscale{0.85}
\plotone{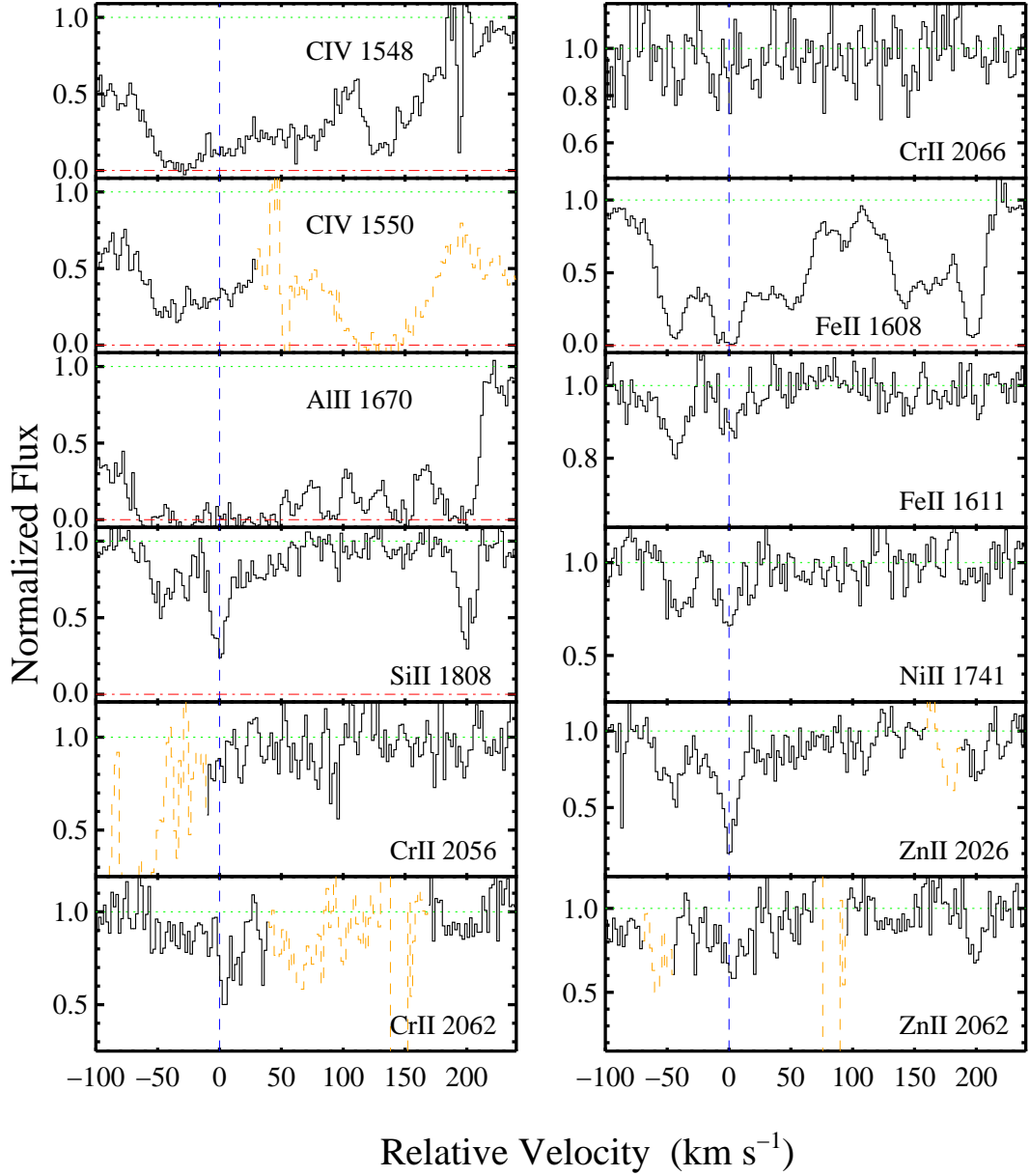}
\caption{HIRES velocity profiles of the transitions identified
with the damped \lya\ systems at $z=3.1040$ toward B1013+0035.
}
\label{fig:b1013_z310}
\end{figure}

\begin{figure}
\epsscale{0.85}
\plotone{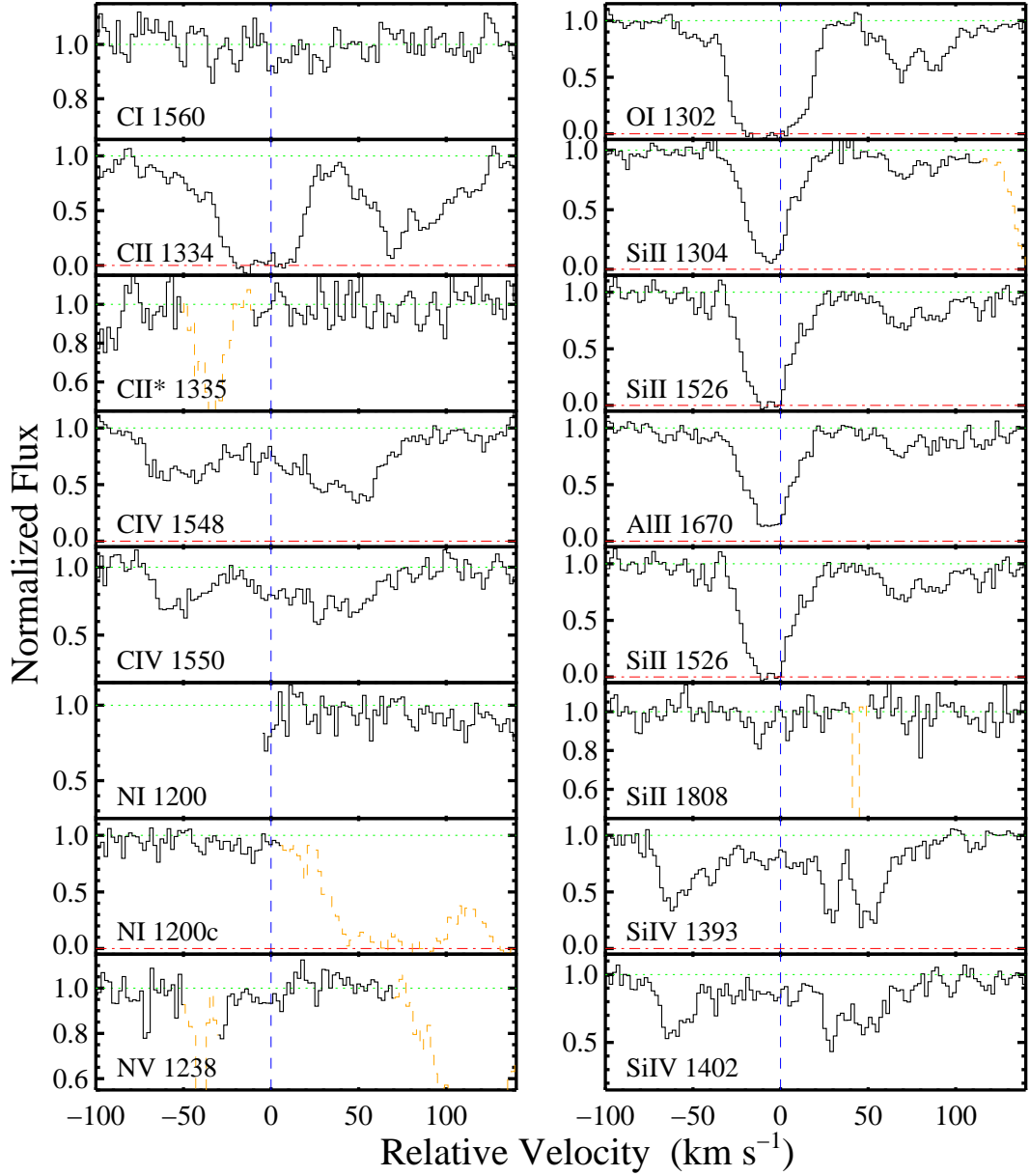}
\caption{HIRES velocity profiles of the transitions identified
with the damped \lya\ systems at $z=2.9489$ toward Q1021+30.
}
\label{fig:q1021_z294}
\end{figure}

\begin{figure}
\epsscale{0.85}
\plotone{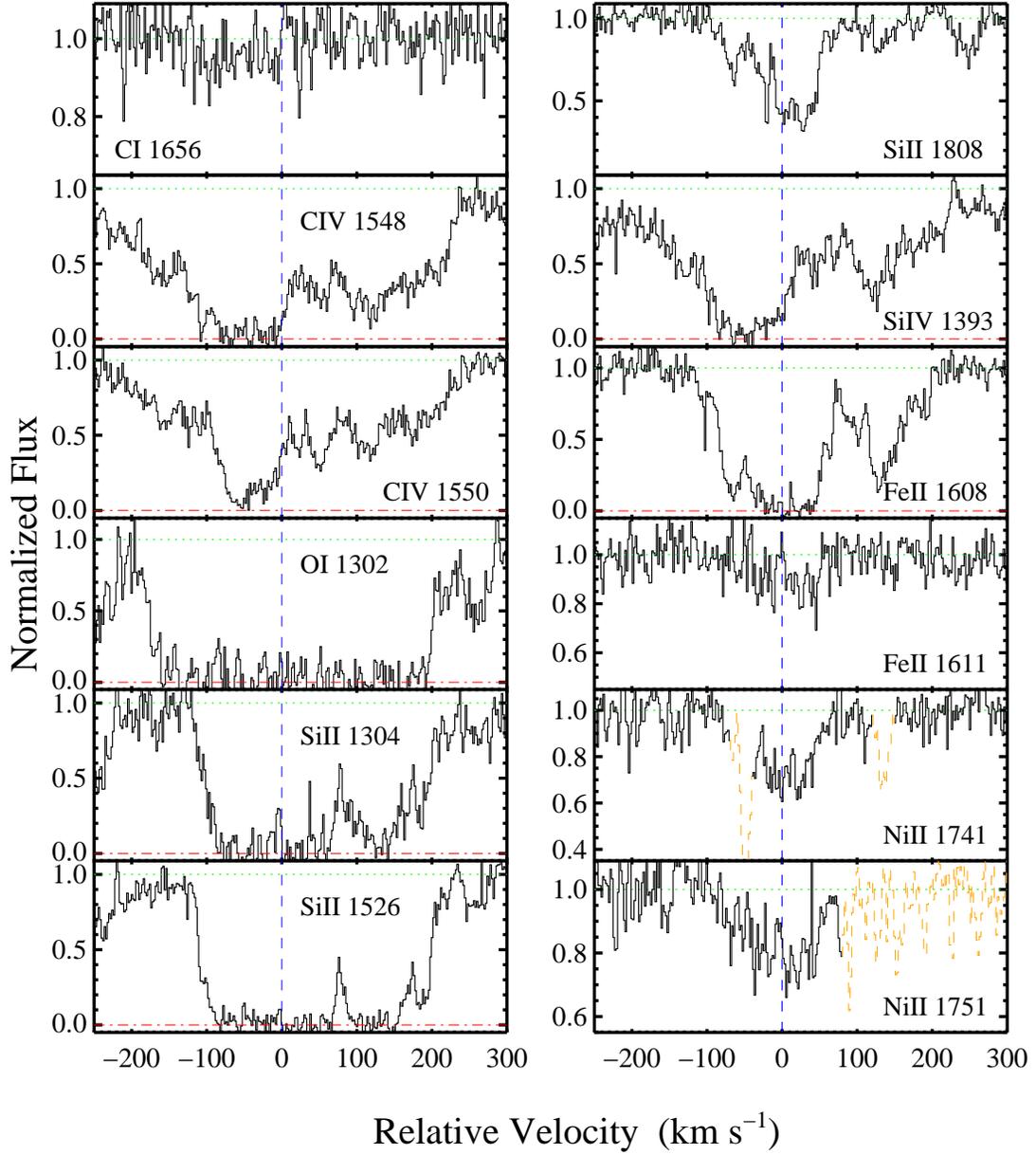}
\caption{HIRES velocity profiles of the transitions identified
with the damped \lya\ systems at $z=2.5841$ toward Q1209+0919.
}
\label{fig:q1209_z258}
\end{figure}

\begin{figure}
\epsscale{0.85}
\plotone{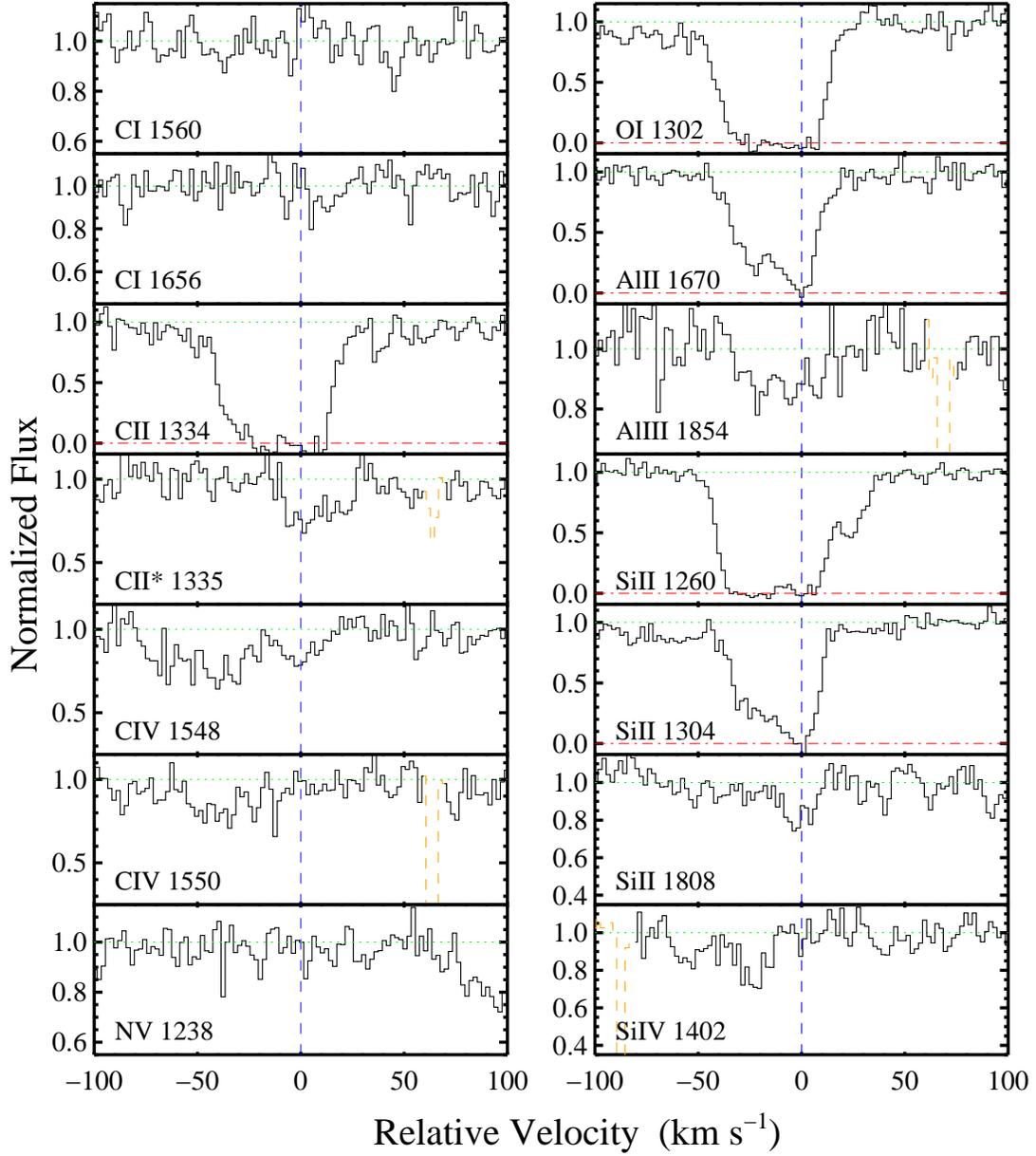}
\caption{HIRES velocity profiles of the transitions identified
with the damped \lya\ systems at $z=2.79585$ toward Q1337+11.
}
\label{fig:q1337_z279}
\end{figure}

\begin{figure}
\epsscale{0.85}
\plotone{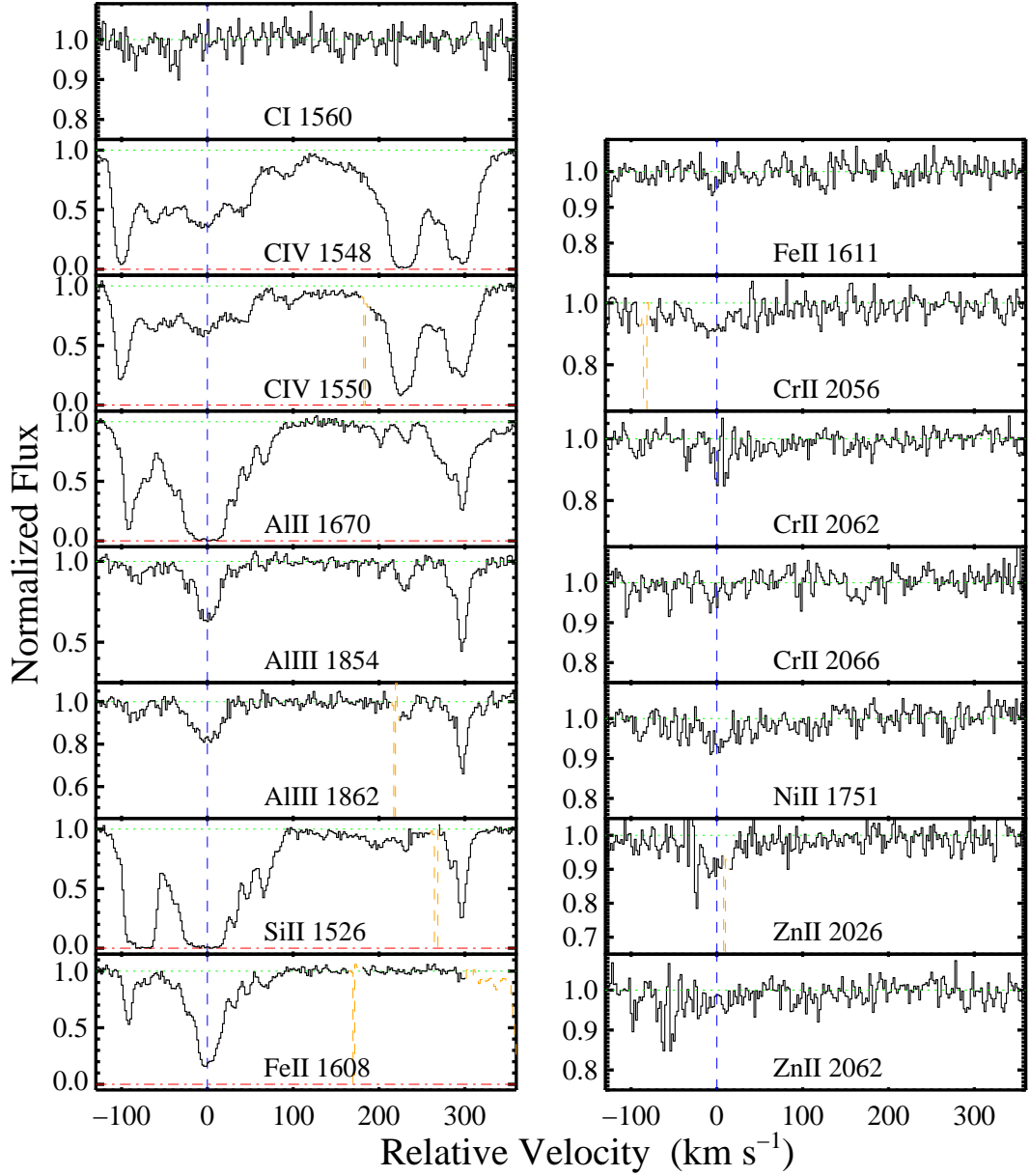}
\caption{HIRES velocity profiles of the transitions identified
with the damped \lya\ systems at $z=2.8268$ toward Q1425+60.
}
\label{fig:q1425_z282}
\end{figure}

\begin{figure}
\epsscale{0.85}
\plotone{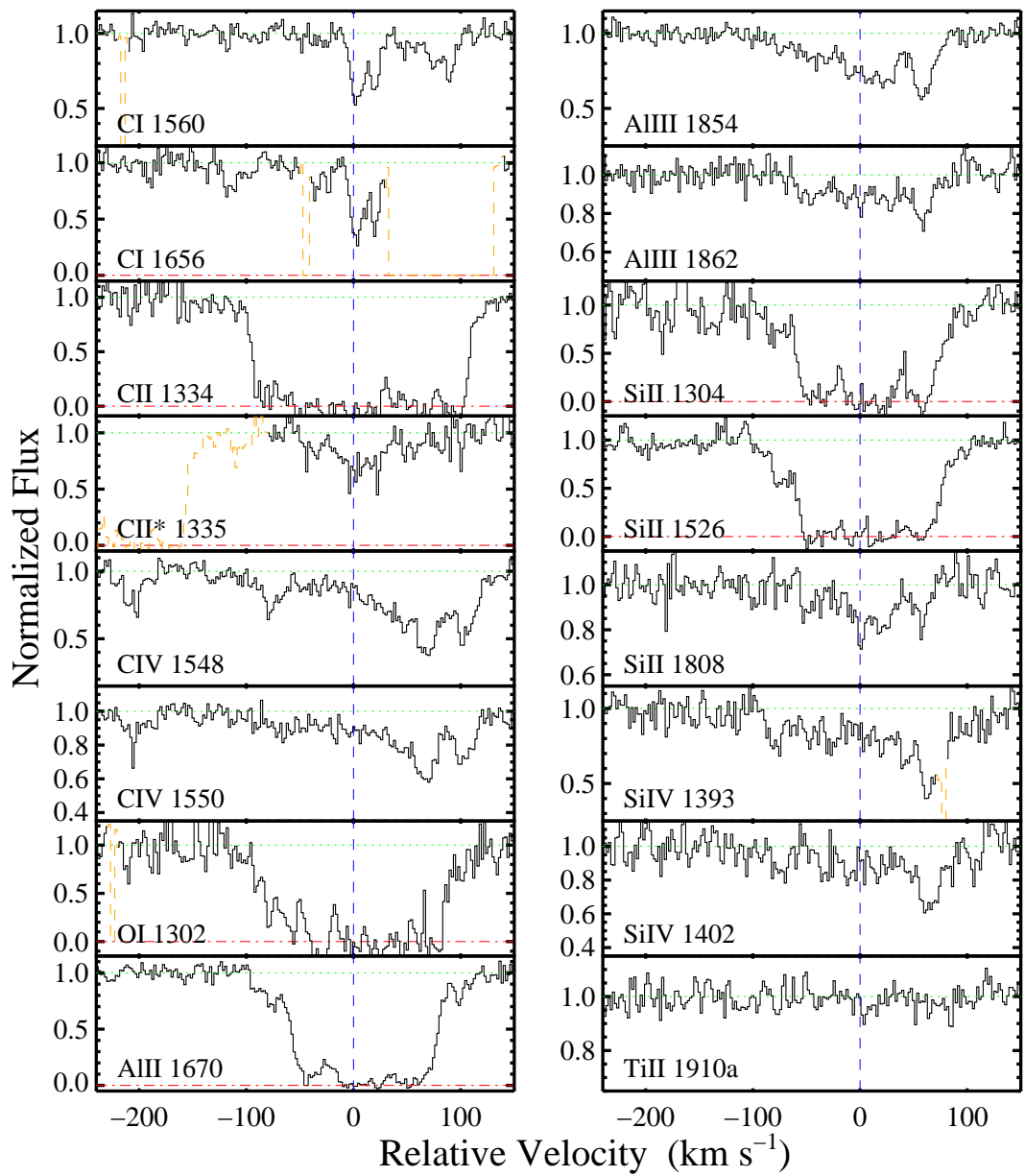}
\caption{HIRES velocity profiles of the transitions identified
with the damped \lya\ systems at $z=2.05452$ toward J2340$-$00.
}
\label{fig:j2340_z205}
\end{figure}

\begin{figure}
\epsscale{0.85}
\plotone{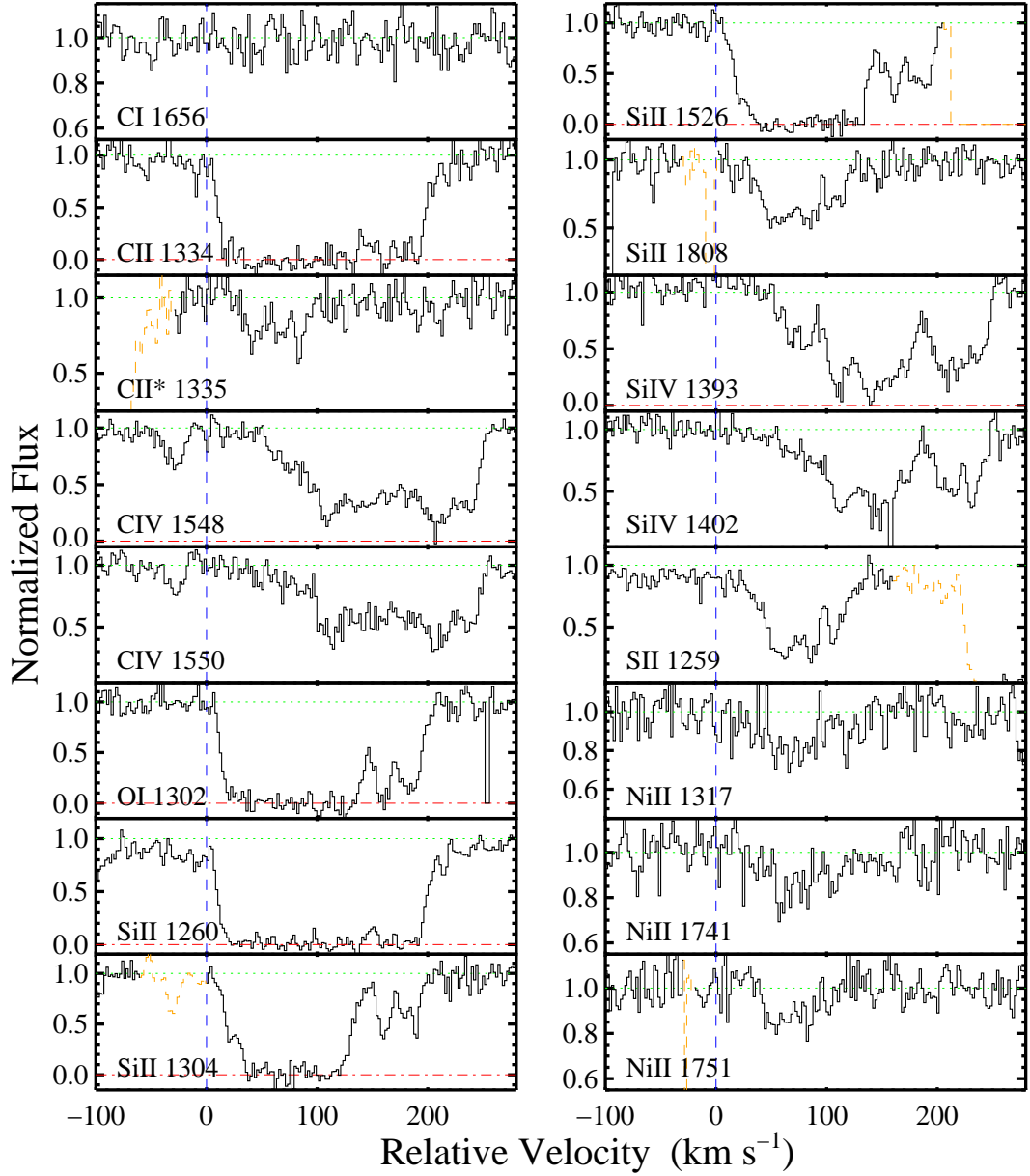}
\caption{HIRES velocity profiles of the transitions identified
with the damped \lya\ systems at $z=2.90823$ toward Q2342+34.
}
\label{fig:q2342_z291}
\end{figure}

Figures~\ref{fig:fj0812_z206}-\ref{fig:q2342_z291} present the HIRES
observations for \newdla\ of the damped \lya\ systems that we had
not previously
published or had published only partially \citep[e.g. FJ0812+32;][]{phw03}.
In each case, blends or bad spectral regions are indicated by dotted lines.
The dashed vertical line at $v=0$\kms\ corresponds to a somewhat arbitrary
redshift (often corresponding to the peak optical depth in low-ion profiles)
given in the figure captions.
The dash-dot line indicates zero flux and the dashed line at unity
traces the normalized quasar continuum.

We have derived ionic column densities from these 
data using primarily the apparent optical depth method \citep{savage91}.
In a few cases (e.g.\ \ion{S}{2} for the DLA at $z=2.626$ toward FJ0812+32), 
we have fitted Voigt profiles to the data using the VPFIT software package
kindly provided by R. Carswell and J. Webb.   In these cases, we adopt
the total column density reported by the package.
Otherwise we adopt the weighted-mean of the observed transitions or
the most stringent upper/lower ($2\sigma$) limit.
All of the measurements are given in 
Tables~\ref{tab:fj0812_z206}-\ref{tab:q2342_z291}.
The errors only reflect statistical uncertainty using standard
error propagation.  These are unrealistic in the case of very high
S/N (where $\sigma \leq 0.01$\,dex) or for very weak transitions
where uncertainty in continuum placement dominates.
We recommend a minimum error of 0.01\,dex for strong transitions
with high S/N data and 0.10\,dex for weak transitions
(peak optical depth less than 10$\%$).
See \cite{pro01} for a full description of our analysis procedures and
a list of the atomic data, drawn primarily from \cite{morton03}.



In those cases where we have obtained both ESI and HIRES observations
on a given sightline (e.g.\ Q1021+30, Q1209+09, Q1337+11, Q2342+34), we report 
the combined abundance measurements.  With few exceptions,
we give the HIRES observations precedence for the transitions
where both datasets provide measurements.
If our group had previously published column density measurements
\citep[e.g.\ Q1425+60;][]{pro01}, 
the full list is presented.

\section{DLA METAL SUMMARY}
\label{sec:summ}

We now summarize the current set of high-resolution
damped \lya\ observations by listing all of the published metallicity
measurements as of September 2006 (Table~\ref{tab:mtl}).
We restrict the summary to optical data (i.e.\ $z_{abs} > 1.6$)
acquired at spectral resolution $R>5500$, i.e.\
echelle or echellette observations.
Although lower resolution observations can give accurate metal
abundance measurements 
for favorable optical depth profiles
\citep[e.g.][]{pettini94,kkl04}, 
we prefer to restrict the list to cases where 
the optical depth profile is at least partially resolved.

In general, the metallicities listed in Table~\ref{tab:mtl}
are [O/H], [Si/H], [S/H], or [Zn/H], in that order of
preference.  The elements O, S, and Zn have the advantage
of being non-refractory and therefore the gas-phase abundances
should reflect the total abundance.  The difficulty with the latter
element (Zn), however, is that it is a trace element
(i.e.\ one expects 1 Zn atom per 10,000 O atoms) and it
has an uncertain nucleosynthetic origin \citep{hwf+96}.
It is difficult to measure O because its transitions are
generally saturated while S often is lost within the \lya\
forest or not covered by our observations.  As such, Si is 
not frequently used for the metallicity.

\begin{figure}
\epsscale{0.85}
\plotone{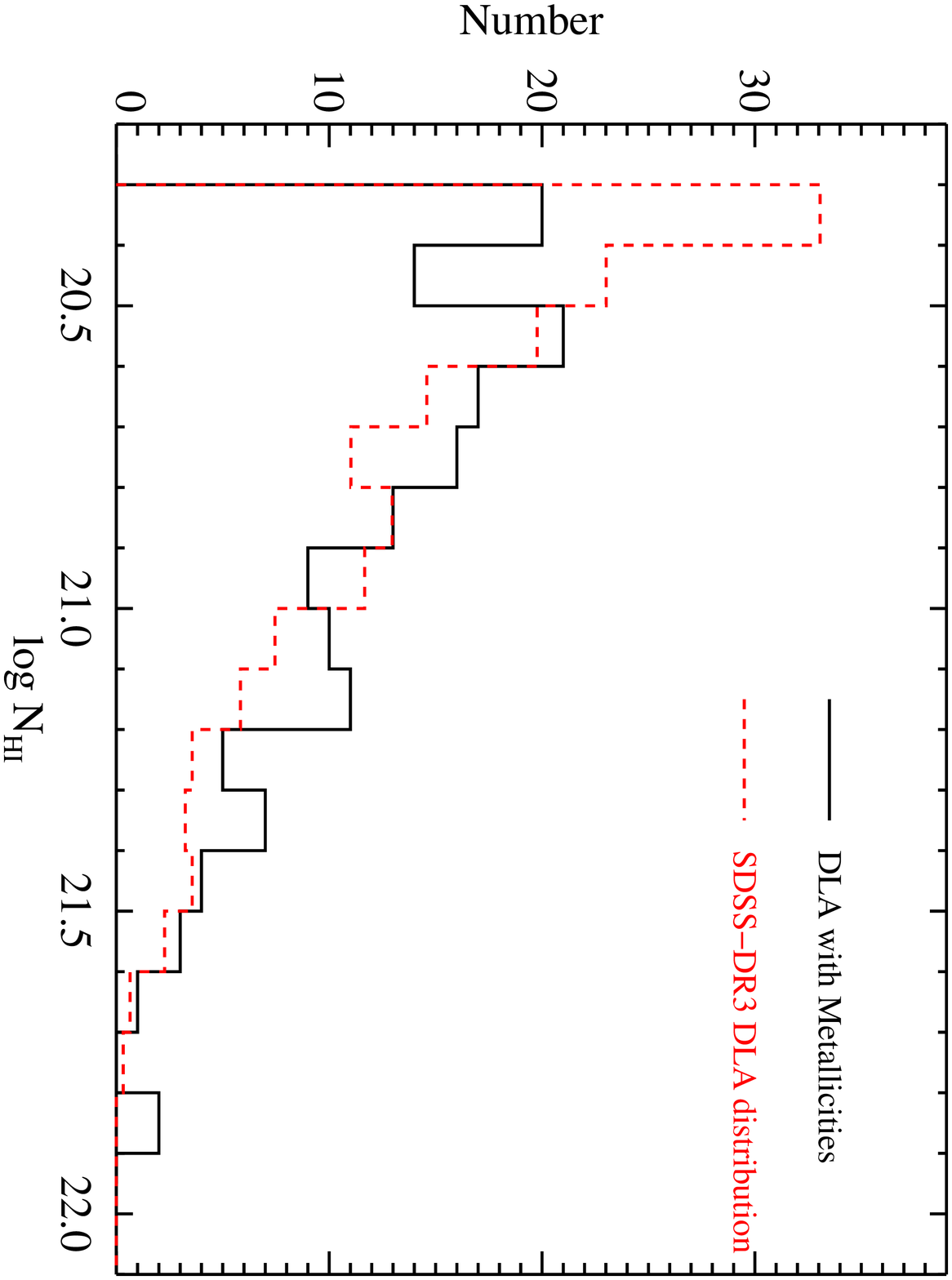}
\caption{\nhi\ histogram for the damped \lya\ systems comprising
the $z>1.6$, high-resolution sample of metallicity measurements
(Table~\ref{tab:mtl}).
Overplotted on the histogram is the expected \nhi\ distribution
for a random set of damped \lya\ systems with the same number
as our metallicity sample.  This curve was 
generated from the \nhi\ frequency distribution measured by
\cite{phw05} from the Sloan Digital Sky Survey.   A two-sided
KS test rules out the null hypothesis that the two distributions
are drawn from the same parent population at $>99\%$c.l.
}
\label{fig:nhi}
\end{figure}

Although Table~\ref{tab:mtl} is a reasonably complete list of 
observed DLA systems, we emphasize that it should not be considered
a representative sample of DLA systems.  In particular, the
\nhi\ frequency distribution of the sample does not follow
that of a `random' set of DLA systems. 
Figure~\ref{fig:nhi} presents a histogram of the \nhi\ values
for the metallicity sample given by Table~\ref{tab:mtl}.
Overplotted on the histogram is the expected \nhi\ distribution
for a random set of damped \lya\ systems.  This curve was 
generated from the \nhi\ frequency distribution measured by
\cite{phw05} from the Sloan Digital Sky Survey, Data Release 3
\cite{sdssdr3}.   
It is evident that the metallicity sample has too few DLA
systems with $\mnhi < 10^{20.6} \cm{-2}$. 
A two-sided Kolmogorov-Smirnov test 
rules out the null hypothesis that the two distributions
are drawn from the same parent population at $>99\%$c.l.

We believe the difference shown in Figure~\ref{fig:nhi}
results primarily because observers
(like ourselves) intentionally avoided low \nhi\ DLA systems
when obtaining high resolution observations to be certain that
their sample satisfied the $\mnhi \geq 2 \sci{20} \cm{-2}$ criterion.   
Furthermore,
we expect that some studies \citep[e.g.\ H$_2$;][]{ledoux03} focused
on large \nhi\ absorbers to increase their likelihood of detections.
Because DLA metallicity is not strongly correlated with \nhi\
\citep{pgw+03}, the metallicity sample in Table~\ref{tab:mtl}
may not be significantly biased. 
Nevertheless, one must take care when drawing conclusions
from this and other DLA datasets.

\section{CONCLUDING REMARKS}
\label{sec:conclude}

In the future, our HIRES observations will be publically available
via the Keck Observatory Archive.  We also plan to release
fully reduced, normalized 1D spectra in a fashion similar to this paper.
We also plan to publish the next set of ESI spectra within the
next two years.

\acknowledgments

The authors wish to recognize and acknowledge the very significant
cultural role and reverence that the summit of Mauna Kea has always
had within the indigenous Hawaiian community.  We are most fortunate
to have the opportunity to conduct observations from this mountain.
JXP is partially supported by NSF grant AST 05-48180 and
JXP and AMW acknowledge support from NSF grant AST 03-07408.
This material is based upon work supported by the National Science 
Foundation under grant AST-0201667, an NSF Astronomy and Astrophysics 
Postdoctoral Fellowship (AAPF) awarded to E. Gawiser.  



\begin{deluxetable}{lccccccccl}
\rotate
\tablewidth{0pc}
\tablecaption{JOURNAL OF HIRES OBSERVATIONS\label{tab:hires}}
\tabletypesize{\footnotesize}
\tablehead{
\colhead{QSO} & \colhead{Alt. Name} & \colhead{RA} & \colhead{DEC} & \colhead{Date (UT)} & \colhead{Exp}
& \colhead{Wavelengths} & \colhead{FWHM} & \colhead{S/N$^a$} & \colhead{Conditions} \\
& & (J2000) & (J2000) & & (s) & (\AA) & (\kms) & (pix$^{-1}$) & }
\startdata
Q0000$-$26 &                 & 00:03:22.9 & $-26$:03:17 & 30 Sep 1994 & 10800 & 6310--8130 & 8 & 15 & \\
           &                 &            &             & 01 Oct 1994 &  3600 & 6410--8130 & 8 &    & \\
Q0019$-$15 & BR 0019$-$1522  & 00:22:08.0 & $-15$:05:39 & 20 Sep 1996 & 13500 & 5300--7800 & 8 & 18 & Clouds \\
           &                 &            &             & 21 Sep 1996 & 12600 & 5890--8320 & 6 &    & Clouds \\
           &                 &            &             & 22 Sep 1996 & 14400 & 5890--8320 & 8 &    & Clouds \\
Q0100+13   & PHL 957         & 01:03:11.4 & +13:16:17   & Spring 1994 & 11700 & 4420--6890 & 6 & 40  \\
Q0149+33   & OC 383          & 01:52:34.4 & +33:50:33   & 29 Sep 1997 & 8100  & 4080--6520 & 8 & 25 & \\
           &                 &            &             & 30 Sep 1997 & 9500  & 4080--6520 & 8 &    & \\
\enddata
\tablenotetext{a}{Total S/N ratio of the final spectrum at the central wavelength of the data.}
\end{deluxetable}

\begin{deluxetable}{lccccccc}
\rotate
\tablewidth{0pc}
\tablecaption{JOURNAL OF ESI OBSERVATIONS\label{tab:esi}}
\tabletypesize{\footnotesize}
\tablehead{
\colhead{QSO} & \colhead{RA} & \colhead{DEC} & \colhead{$z_{em}$} & \colhead{Date (UT)} 
& \colhead{FWHM} & \colhead{Exp} & \colhead{S/N$^a$} \\
& (J2000) & (J2000) & &     & (\kms) & (s) & (pix$^{-1}$) }
\startdata
PX0034+16     & 00 34 54.8 & $+16$ 39 20 & 4.29 & 15 Oct 2001 & 33  & 4200 & 27 \\
PSS0106+2601  & 01 06 00.8 & $+26$ 01 03 & 4.32 & 22 Jan 2001 & 44 & 1800 & 7 \\
Q0112-30      & 01 15 04.8 & $-30$ 25 14 & 2.99 & 15 Oct 2001 & 33  & 2400 & 6 \\
SDSS0127$-$00& 01 27 00.7 & $-00$ 45 59 & 4.06 & 22 Jan 2001 & 44 & 1800 & 23 \\
               &            &             &      & 19 Jul 2001 & 33 & 1800 \\
PSS0133+0400   & 01 33 40.4 & $+04$ 00 59 & 4.13 & 22 Jan 2001 & 44 & 1800 & 28 \\
PSS0134+3317   & 01 34 21.6 & $+33$ 07 56 & 4.52 & 15 Oct 2001 & 33  & 6000 & 25\\
PSS0209+0517   & 02 09 44.7 & $+05$ 17 14 & 4.18 & 22 Jan 2001 & 44 & 1200 & 29\\
\enddata
\tablenotetext{a}{Median signal-to-noise per 11\kms\ pixel at $\lambda \approx 7100$\AA.}
\tablecomments{All of the observing nights were reasonably clear with seeing FWHM~$<1''$.} 
\end{deluxetable}

\clearpage
\begin{deluxetable}{lccccccc}
\tablewidth{0pc}
\tablecaption{IONIC COLUMN DENSITIES FOR FJ0812+32 $z=2.0668$\label{tab:fj0812_z206}}
\tabletypesize{\footnotesize}
\tablehead{\colhead{Ion} & \colhead{$\lambda$} & \colhead{$\log f$}
& \colhead{Instr.} 
& \colhead{$v_{int}^a$} 
& \colhead{$W_\lambda^b$} 
& \colhead{$\log N$} & \colhead{$\log N_{adopt}$} \\
& (\AA) & & & (\kms) & (m\AA) & & }
\startdata
\ion{C}{1}\\
&1560.3092 &$ -0.8808$&HIRES&$[ -15,   5]$&$   7.7 \pm   1.1$&$ 12.47 \pm 0.06$&$ 12.60 \pm 0.02$\\
&1656.9283 &$ -0.8273$&HIRES&$[ -15,   2]$&$  13.2 \pm   0.8$&$ 12.64 \pm 0.03$& \\
\ion{C}{4}\\
&1548.1950 &$ -0.7194$&HIRES&$[-100, 100]$&$ 252.0 \pm   2.6$&$ 13.95 \pm 0.01$&$ 13.95 \pm 0.01$\\
&1550.7700 &$ -1.0213$&HIRES&$[-100, 100]$&$ 148.0 \pm   2.9$&$ 13.95 \pm 0.01$& \\
\ion{Al}{2}\\
&1670.7874 &$  0.2742$&HIRES&$[ -80,  40]$&$ 220.0 \pm   2.2$&$> 13.13$&$> 13.13$\\
\enddata
\tablenotetext{a}{Velocity interval over which the equivalent width and
column density are measured.}
\tablenotetext{b}{Rest equivalent width.}
\end{deluxetable}

\clearpage
\begin{deluxetable}{lccccccc}
\tablewidth{0pc}
\tablecaption{IONIC COLUMN DENSITIES FOR FJ0812+32 $z=2.6263$\label{tab:fj0812_z262}}
\tabletypesize{\footnotesize}
\tablehead{\colhead{Ion} & \colhead{$\lambda$} & \colhead{$\log f$}
& \colhead{Instr.} 
& \colhead{$v_{int}^a$} 
& \colhead{$W_\lambda^b$} 
& \colhead{$\log N$} & \colhead{$\log N_{adopt}$} \\
& (\AA) & & & (\kms) & (m\AA) & & }
\startdata
\ion{B}{2}\\
&1362.4610 &$ -0.0057$&HIRES&$[ -15,  15]$&$   4.3 \pm   0.9$&$ 11.43 \pm 0.09$&$ 11.43 \pm 0.09$\\
\ion{C}{1}\\
&1328.8333 &$ -1.0467$&HIRES&$[ -20,  20]$&$  10.5 \pm   1.1$&$ 12.90 \pm 0.04$&$ 12.75 \pm 0.03$\\
&1560.3092 &$ -0.8808$&HIRES&$[ -20,  20]$&$  13.7 \pm   1.0$&$ 12.71 \pm 0.03$& \\
&1656.9283 &$ -0.8273$&HIRES&$[ -20,  20]$&$  19.4 \pm   1.1$&$> 12.77$& \\
\ion{C}{2}\\
&1334.5323 &$ -0.8935$&HIRES&$[-110,  70]$&$ 641.1 \pm   1.7$&$> 15.20$&$> 15.20$\\
\enddata
\tablenotetext{a}{Velocity interval over which the equivalent width and
column density are measured.}
\tablenotetext{b}{Rest equivalent width.}
\end{deluxetable}

\clearpage
\begin{deluxetable}{lccccccc}
\tablewidth{0pc}
\tablecaption{IONIC COLUMN DENSITIES FOR J0900+42 $z=3.2458$\label{tab:j0900_z324}}
\tabletypesize{\footnotesize}
\tablehead{\colhead{Ion} & \colhead{$\lambda$} & \colhead{$\log f$}
& \colhead{Instr.} 
& \colhead{$v_{int}^a$} 
& \colhead{$W_\lambda^b$} 
& \colhead{$\log N$} & \colhead{$\log N_{adopt}$} \\
& (\AA) & & & (\kms) & (m\AA) & & }
\startdata
\ion{C}{1}\\
&1328.8333 &$ -1.0467$&HIRES&$[ -20,  20]$&$<   1.4$&$< 12.18$&$ 11.79 \pm 0.14$\\
&1560.3092 &$ -0.8808$&HIRES&$[ -20,  20]$&$<   1.1$&$< 11.75$& \\
&1656.9283 &$ -0.8273$&HIRES&$[  -5,  20]$&$   2.2 \pm   0.7$&$ 11.79 \pm 0.14$& \\
\ion{C}{2}\\
&1334.5323 &$ -0.8935$&HIRES&$[-100, 100]$&$ 624.6 \pm   1.1$&$> 15.23$&$> 15.23$\\
&1335.7077 &$ -0.9397$&HIRES&$[ -70,  20]$&$  15.0 \pm   1.0$&$ 12.93 \pm 0.03$& \\
\ion{C}{4}\\
\enddata
\tablenotetext{a}{Velocity interval over which the equivalent width and
column density are measured.}
\tablenotetext{b}{Rest equivalent width.}
\end{deluxetable}

\clearpage
\begin{deluxetable}{lccccccc}
\tablewidth{0pc}
\tablecaption{IONIC COLUMN DENSITIES FOR B1013+0035 $z=3.1040$\label{tab:b1013_z310}}
\tabletypesize{\footnotesize}
\tablehead{\colhead{Ion} & \colhead{$\lambda$} & \colhead{$\log f$}
& \colhead{Instr.} 
& \colhead{$v_{int}^a$} 
& \colhead{$W_\lambda^b$} 
& \colhead{$\log N$} & \colhead{$\log N_{adopt}$} \\
& (\AA) & & & (\kms) & (m\AA) & & }
\startdata
\ion{C}{4}\\
&1548.1950 &$ -0.7194$&HIRES&$[-100, 200]$&$1068.2 \pm   7.1$&$> 14.77$&$> 14.77$\\
\ion{Al}{2}\\
&1670.7874 &$  0.2742$&HIRES&$[-100, 240]$&$1592.6 \pm   8.4$&$> 14.07$&$> 14.07$\\
\ion{Si}{2}\\
&1808.0130 &$ -2.6603$&HIRES&$[ -90, 230]$&$ 295.6 \pm  10.5$&$ 15.78 \pm 0.02$&$ 15.78 \pm 0.02$\\
\ion{Cr}{2}\\
&2056.2539 &$ -0.9788$&HIRES&$[   0, 100]$&$  54.5 \pm  10.7$&$> 13.21$&$> 13.21$\\
&2066.1610 &$ -1.2882$&HIRES&$[ -80, 120]$&$<  29.6$&$< 13.45$& \\
\ion{Fe}{2}\\
&1608.4511 &$ -1.2366$&HIRES&$[-100, 240]$&$ 883.1 \pm   4.0$&$> 15.15$&$ 15.18 \pm 0.05$\\
&1611.2005 &$ -2.8665$&HIRES&$[-100, 240]$&$  43.1 \pm   5.1$&$ 15.18 \pm 0.05$& \\
\ion{Ni}{2}\\
&1741.5531 &$ -1.3696$&HIRES&$[ -80, 200]$&$  71.8 \pm  10.1$&$ 13.89 \pm 0.05$&$ 13.89 \pm 0.05$\\
\ion{Zn}{2}\\
&2026.1360 &$ -0.3107$&HIRES&$[ -80, 240]$&$ 298.4 \pm  12.7$&$ 13.33 \pm 0.02$&$ 13.33 \pm 0.02$\\
\enddata
\tablenotetext{a}{Velocity interval over which the equivalent width and
column density are measured.}
\tablenotetext{b}{Rest equivalent width.}
\end{deluxetable}

\clearpage
\begin{deluxetable}{lccccccc}
\tablewidth{0pc}
\tablecaption{IONIC COLUMN DENSITIES FOR Q1021+30 $z=2.9489$\label{tab:q1021_z295}}
\tabletypesize{\footnotesize}
\tablehead{\colhead{Ion} & \colhead{$\lambda$} & \colhead{$\log f$}
& \colhead{Instr.} 
& \colhead{$v_{int}^a$} 
& \colhead{$W_\lambda^b$} 
& \colhead{$\log N$} & \colhead{$\log N_{adopt}$} \\
& (\AA) & & & (\kms) & (m\AA) & & }
\startdata
\ion{C}{1}\\
&1560.3092 &$ -0.8808$&HIRES&$[ -40,  60]$&$  13.7 \pm   4.2$&$< 12.71$&$< 12.71$\\
&1656.9283 &$ -0.8273$&ESI&$[ -50,  50]$&$<  15.5$&$< 12.81$& \\
\ion{C}{2}\\
&1334.5323 &$ -0.8935$&HIRES&$[-100, 150]$&$ 467.5 \pm   6.4$&$> 14.77$&$> 14.77$\\
&1335.7077 &$ -0.9397$&HIRES&$[ -15,  50]$&$<   7.8$&$< 12.81$& \\
\ion{C}{4}\\
&1548.1950 &$ -0.7194$&HIRES&$[-100, 150]$&$ 316.3 \pm   7.1$&$ 14.01 \pm 0.01$&$ 13.99 \pm 0.01$\\
\enddata
\tablenotetext{a}{Velocity interval over which the equivalent width and
column density are measured.}
\tablenotetext{b}{Rest equivalent width.}
\end{deluxetable}

\clearpage
\begin{deluxetable}{lccccccc}
\tablewidth{0pc}
\tablecaption{IONIC COLUMN DENSITIES FOR Q1209+0919 $z=2.5841$\label{tab:q1209_z258}}
\tabletypesize{\footnotesize}
\tablehead{\colhead{Ion} & \colhead{$\lambda$} & \colhead{$\log f$}
& \colhead{Instr.} 
& \colhead{$v_{int}^a$} 
& \colhead{$W_\lambda^b$} 
& \colhead{$\log N$} & \colhead{$\log N_{adopt}$} \\
& (\AA) & & & (\kms) & (m\AA) & & }
\startdata
\ion{C}{1}\\
&1560.3092 &$ -0.8808$&HIRES&$[ -50,  50]$&$<   8.8$&$< 12.67$&$ 12.60 \pm 0.13$\\
&1656.9283 &$ -0.8273$&HIRES&$[ -60,  60]$&$<   8.6$&$ 12.60 \pm 0.13$& \\
\ion{C}{4}\\
&1548.1950 &$ -0.7194$&HIRES&$[-300, 300]$&$1670.4 \pm  10.4$&$> 14.97$&$> 15.02$\\
&1550.7700 &$ -1.0213$&HIRES&$[-240, 300]$&$1189.4 \pm   9.4$&$> 15.02$& \\
\ion{Al}{2}\\
&1670.7874 &$  0.2742$&ESI&$[-300, 300]$&$1902.7 \pm  23.6$&$> 14.07$&$> 14.07$\\
\enddata
\tablenotetext{a}{Velocity interval over which the equivalent width and
column density are measured.}
\tablenotetext{b}{Rest equivalent width.}
\end{deluxetable}

\clearpage
\begin{deluxetable}{lccccccc}
\tablewidth{0pc}
\tablecaption{IONIC COLUMN DENSITIES FOR Q1337+11 $z=2.79585$\label{tab:q1337_z279}}
\tabletypesize{\footnotesize}
\tablehead{\colhead{Ion} & \colhead{$\lambda$} & \colhead{$\log f$}
& \colhead{Instr.} 
& \colhead{$v_{int}^a$} 
& \colhead{$W_\lambda^b$} 
& \colhead{$\log N$} & \colhead{$\log N_{adopt}$} \\
& (\AA) & & & (\kms) & (m\AA) & & }
\startdata
\ion{C}{1}\\
&1560.3092 &$ -0.8808$&HIRES&$[ -60,  60]$&$<  12.0$&$< 12.81$&$< 12.66$\\
&1656.9284 &$ -0.8273$&HIRES&$[ -40,  40]$&$<  10.8$&$< 12.66$& \\
\ion{C}{2}\\
&1334.5323 &$ -0.8935$&HIRES&$[ -70,  50]$&$ 282.0 \pm   6.1$&$> 14.64$&$> 14.64$\\
&1335.7077 &$ -0.9397$&HIRES&$[ -20,  40]$&$  29.8 \pm   4.4$&$ 13.28 \pm 0.06$& \\
\ion{C}{4}\\
&1548.1950 &$ -0.7194$&HIRES&$[-100,  30]$&$  81.3 \pm   6.0$&$ 13.36 \pm 0.03$&$ 13.39 \pm 0.03$\\
\enddata
\tablenotetext{a}{Velocity interval over which the equivalent width and
column density are measured.}
\tablenotetext{b}{Rest equivalent width.}
\end{deluxetable}

\clearpage
\begin{deluxetable}{lccccccc}
\tablewidth{0pc}
\tablecaption{IONIC COLUMN DENSITIES FOR Q1425+60 $z=2.8268$\label{tab:q1425_z282}}
\tabletypesize{\footnotesize}
\tablehead{\colhead{Ion} & \colhead{$\lambda$} & \colhead{$\log f$}
& \colhead{Instr.} 
& \colhead{$v_{int}^a$} 
& \colhead{$W_\lambda^b$} 
& \colhead{$\log N$} & \colhead{$\log N_{adopt}$} \\
& (\AA) & & & (\kms) & (m\AA) & & }
\startdata
\ion{C}{1}\\
&1560.3092 &$ -0.8808$&ESI&$[ -80,  60]$&$<   4.6$&$< 12.39$&$< 12.39$\\
\ion{C}{2}\\
&1036.3367 &$ -0.9097$&HIRES&$[-120, 120]$&$ 680.2 \pm  18.4$&$> 15.22$&$> 15.23$\\
&1335.7077 &$ -0.9397$&HIRES&$[ -20,  20]$&$  43.4 \pm   0.8$&$< 13.52$& \\
\ion{C}{4}\\
&1548.1950 &$ -0.7194$&ESI&$[-150, 350]$&$1108.0 \pm   3.2$&$> 14.79$&$ 14.73 \pm 0.01$\\
&1550.7700 &$ -1.0213$&ESI&$[-120, 350]$&$ 739.1 \pm  10.9$&$ 14.73 \pm 0.01$& \\
\enddata
\tablenotetext{a}{Velocity interval over which the equivalent width and
column density are measured.}
\tablenotetext{b}{Rest equivalent width.}
\end{deluxetable}

\clearpage
\begin{deluxetable}{lccccccc}
\tablewidth{0pc}
\tablecaption{IONIC COLUMN DENSITIES FOR J2340-00 $z=2.05452$\label{tab:j2340_z205}}
\tabletypesize{\footnotesize}
\tablehead{\colhead{Ion} & \colhead{$\lambda$} & \colhead{$\log f$}
& \colhead{Instr.} 
& \colhead{$v_{int}^a$} 
& \colhead{$W_\lambda^b$} 
& \colhead{$\log N$} & \colhead{$\log N_{adopt}$} \\
& (\AA) & & & (\kms) & (m\AA) & & }
\startdata
\ion{C}{1}\\
&1560.3092 &$ -0.8808$&HIRES&$[ -50,  40]$&$  64.3 \pm   3.3$&$> 13.44$&$> 13.63$\\
&1656.9283 &$ -0.8273$&HIRES&$[ -40,  30]$&$ 113.5 \pm   4.8$&$> 13.63$& \\
\ion{C}{2}\\
&1334.5323 &$ -0.8935$&HIRES&$[-200, 120]$&$ 913.8 \pm  12.7$&$> 15.20$&$> 15.20$\\
&1335.7077 &$ -0.9397$&HIRES&$[ -80, 100]$&$ 105.4 \pm   9.0$&$ 13.84 \pm 0.04$& \\
\ion{C}{4}\\
&1548.1950 &$ -0.7194$&HIRES&$[-220, 130]$&$ 326.9 \pm   7.9$&$ 14.00 \pm 0.01$&$ 14.02 \pm 0.01$\\
\enddata
\tablenotetext{a}{Velocity interval over which the equivalent width and
column density are measured.}
\tablenotetext{b}{Rest equivalent width.}
\end{deluxetable}

\clearpage
\begin{deluxetable}{lccccccc}
\tablewidth{0pc}
\tablecaption{IONIC COLUMN DENSITIES FOR Q2342+34 $z=2.90823$\label{tab:q2342_z291}}
\tabletypesize{\footnotesize}
\tablehead{\colhead{Ion} & \colhead{$\lambda$} & \colhead{$\log f$}
& \colhead{Instr.} 
& \colhead{$v_{int}^a$} 
& \colhead{$W_\lambda^b$} 
& \colhead{$\log N$} & \colhead{$\log N_{adopt}$} \\
& (\AA) & & & (\kms) & (m\AA) & & }
\startdata
\ion{C}{1}\\
&1560.3092 &$ -0.8808$&HIRES&$[   0, 120]$&$<  15.3$&$< 12.92$&$< 12.68$\\
&1656.9283 &$ -0.8273$&HIRES&$[   0, 100]$&$<  11.3$&$< 12.68$& \\
\ion{C}{2}\\
&1334.5323 &$ -0.8935$&HIRES&$[ -50, 240]$&$ 862.8 \pm  10.9$&$> 15.15$&$> 15.15$\\
&1335.7077 &$ -0.9397$&HIRES&$[   0, 130]$&$  62.9 \pm   8.2$&$ 13.61 \pm 0.06$& \\
\ion{C}{4}\\
&1548.1950 &$ -0.7194$&HIRES&$[ -90, 280]$&$ 662.4 \pm   9.6$&$> 14.43$&$ 14.46 \pm 0.01$\\
\enddata
\tablenotetext{a}{Velocity interval over which the equivalent width and
column density are measured.}
\tablenotetext{b}{Rest equivalent width.}
\end{deluxetable}

\begin{deluxetable}{lccccccccccccccc}
\rotate
\tablewidth{0pc}
\tablecaption{DLA METAL SUMMARY\label{tab:mtl}}
\tabletypesize{\scriptsize}
\tablehead{\colhead{Quasar} &\colhead{RA (2000)} & \colhead{DEC (2000)} & \colhead{$z_{em}$} & \colhead{$z_{abs}$} & \colhead{log \nhi} & 
\colhead{f$_\alpha^a$} & \colhead{[$\alpha$/H]} &
\colhead{$\sigma(\alpha$)} &
\colhead{f$_{Zn}^b$} & \colhead{[Zn/H]} &
\colhead{$\sigma$(Zn)} &
\colhead{f$_{Fe}^c$} & \colhead{[Fe/H]} &
\colhead{$\sigma$(Fe)} & \colhead{Ref}}
\startdata
SDSS1435+0420&14:35:12.94&+04:20:36&1.950&1.6559&$21.25^{+0.15}_{-0.15}$& 1&$-0.89$&0.07& 3&$-0.55$&& 1&$-1.05$&0.07&29\\
Q1104-18&11:06:32.96&-18:21:09.82&2.310&1.6614&$20.80^{+0.10}_{-0.10}$& 1&$-1.04$&0.01& 1&$-0.99$&0.01& 1&$-1.48$&0.02&7\\
Q1331+17&13:33:35.78&+16:49:04.03&2.080&1.7764&$21.14^{+0.08}_{-0.08}$& 1&$-1.42$&0.00& 1&$-1.27$&0.03& 1&$-2.02$&0.00&6,13\\
SDSS1249-0233&12:49:24.86&-02:33:39&2.120&1.7806&$21.45^{+0.15}_{-0.15}$& 1&$-1.21$&0.03& 1&$-1.01$&0.04& 1&$-1.48$&0.02&29\\
B2314-409&23:16:46.94&-40:41:21.3&2.448&1.8573&$20.90^{+0.10}_{-0.10}$& 1&$-1.05$&0.11& 1&$-1.05$&0.11& 1&$-1.32$&0.11&22\\
Q0841+12&08:44:24.289&+12:45:48.74&2.500&1.8640&$21.00^{+0.10}_{-0.10}$& 4&$-1.51$&0.03& 0&$$&& 0&$$&&26\\
Q2230+02&22:32:35.30&+02:47:55.12&2.150&1.8644&$20.85^{+0.08}_{-0.08}$& 1&$-0.75$&0.01& 1&$-0.72$&0.03& 1&$-1.16$&0.02&6,13\\
Q1210+17&12:13:03.07&+17:14:23.01&2.540&1.8918&$20.60^{+0.10}_{-0.10}$& 1&$-0.87$&0.02& 1&$-0.90$&0.03& 1&$-1.14$&0.06&13,30\\
\enddata
\tablenotetext{a}{0=No measurement; 1=Si measurement; 2=Si lower limit;
3=Si upper limit; 4=[S/H] ; 5=[O/H] ; 13=S+Si limits}
\tablenotetext{b}{0=No measurement; 1=Zn measurement; 2=Zn lower limit;
3=Zn upper limit}
\tablenotetext{c}{0=No measurement; 1=Fe measurement; 2=Fe lower limit;
3=Fe upper limit; 4=[Ni/H]-0.1dex;
5=[Cr/H] - 0.2dex;
6=[Al/H]; 11-16=Fe, Ni, Cr, Al limits; 25=Mean of Fe limits.}
\tablecomments{Note that none of the limits reported take into account the
uncertainty in the \nhi\ value.}
\tablerefs{
1: \cite{wolfe94};
2: \cite{lu96};
3: \cite{pw96};
4: \cite{pw97b};
5: \cite{lu99};
6: \cite{pw99};
7: \cite{lopez99};
8: \cite{pw00};
9: \cite{petit00};
10: \cite{molaro00};
11: \cite{mirka01};
12: \cite{molaro01};
13: \cite{pro01};
14: \cite{pgw01};
15: \cite{eps+01};
16: \cite{pho+02};
17: \cite{lsp02};
18: \cite{lrd+02};
19: \cite{ldd+02};
20: \cite{lopez03};
21: \cite{songaila02};
22: \cite{p03_esi};
23: \cite{pro03b};
24: \cite{ledoux03};
25: \cite{mirka04};
26: \cite{ledoux06};
27: \cite{akerman05};
28: \cite{obp+06};
29: \cite{shf06};
30: \cite{dz06};
31: This paper }
 
\end{deluxetable}

\clearpage
\pagestyle{plaintop}

\clearpage
\epsscale{0.85}
{\plotone{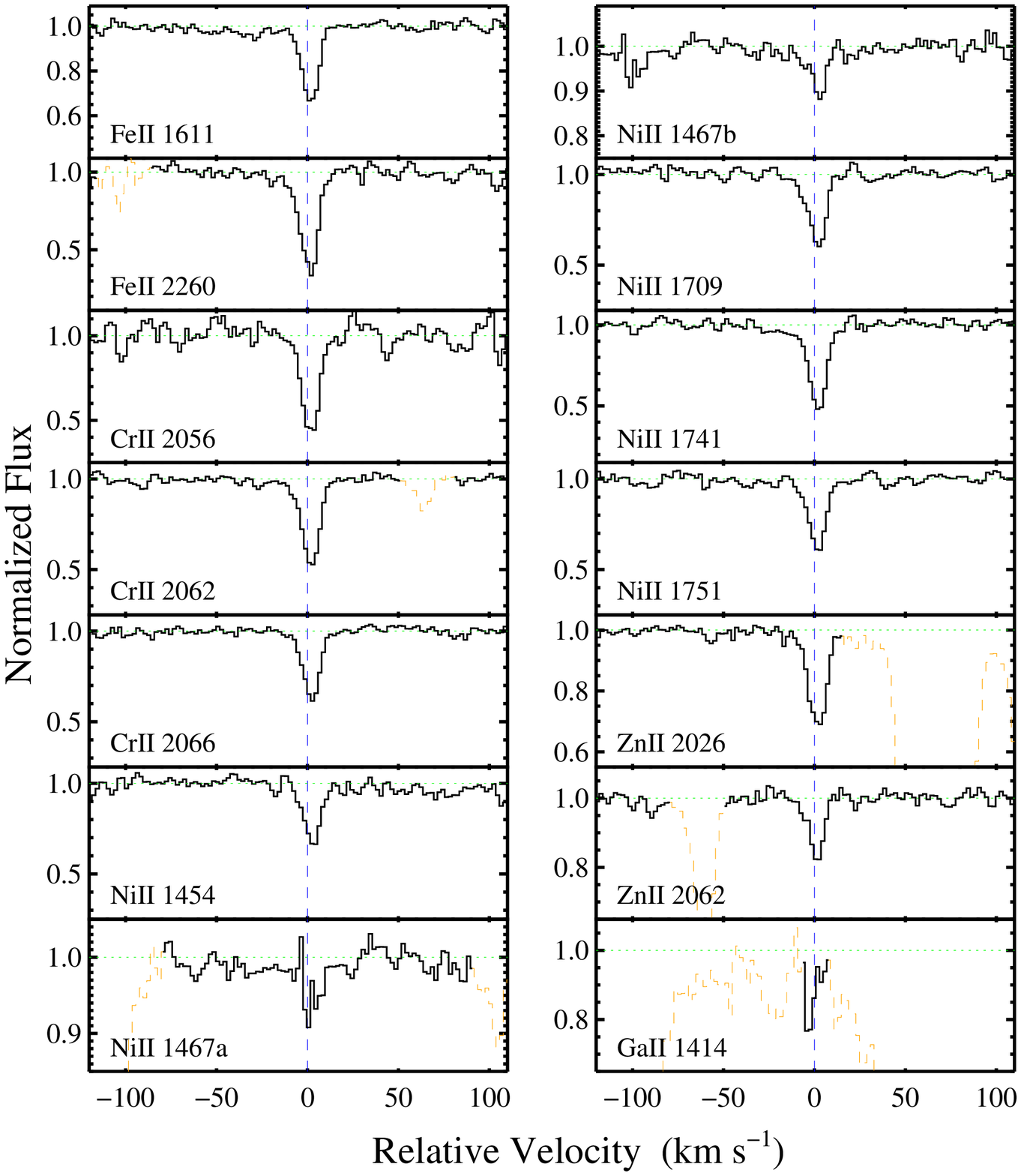}}
\clearpage

\clearpage
\epsscale{0.85}
{\plotone{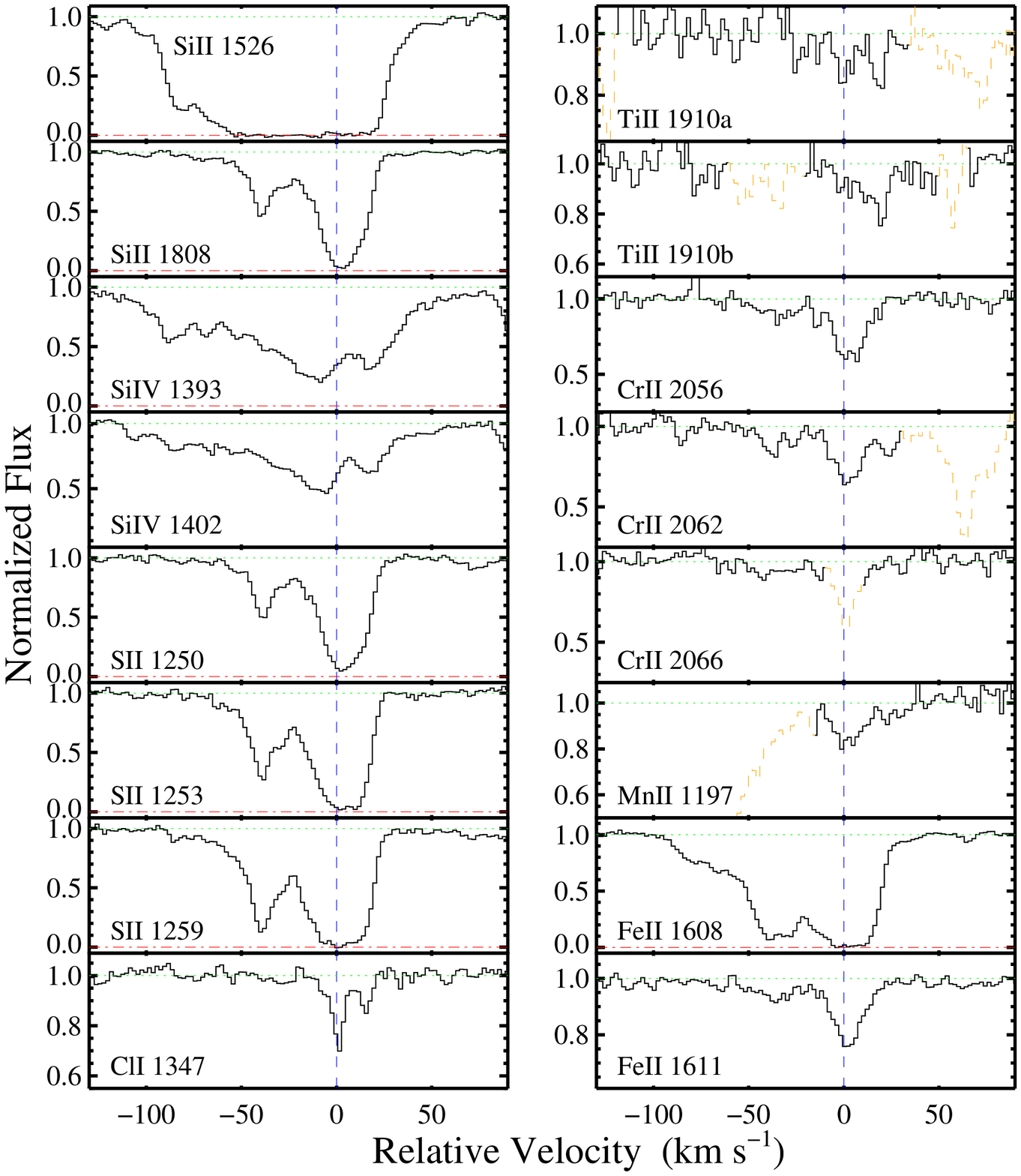}}
\clearpage
{\plotone{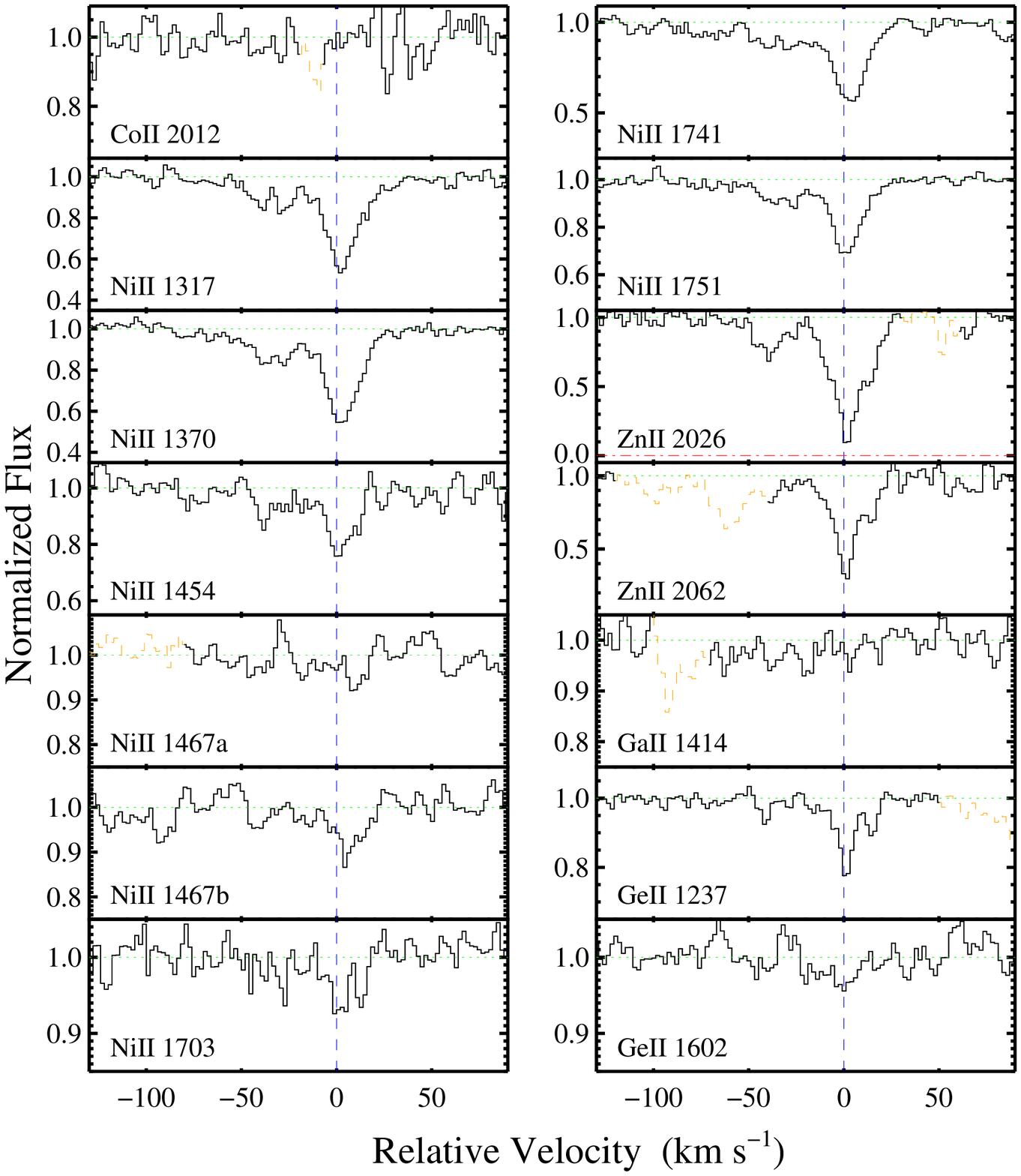}}
\clearpage

\clearpage
\epsscale{0.85}
{\plotone{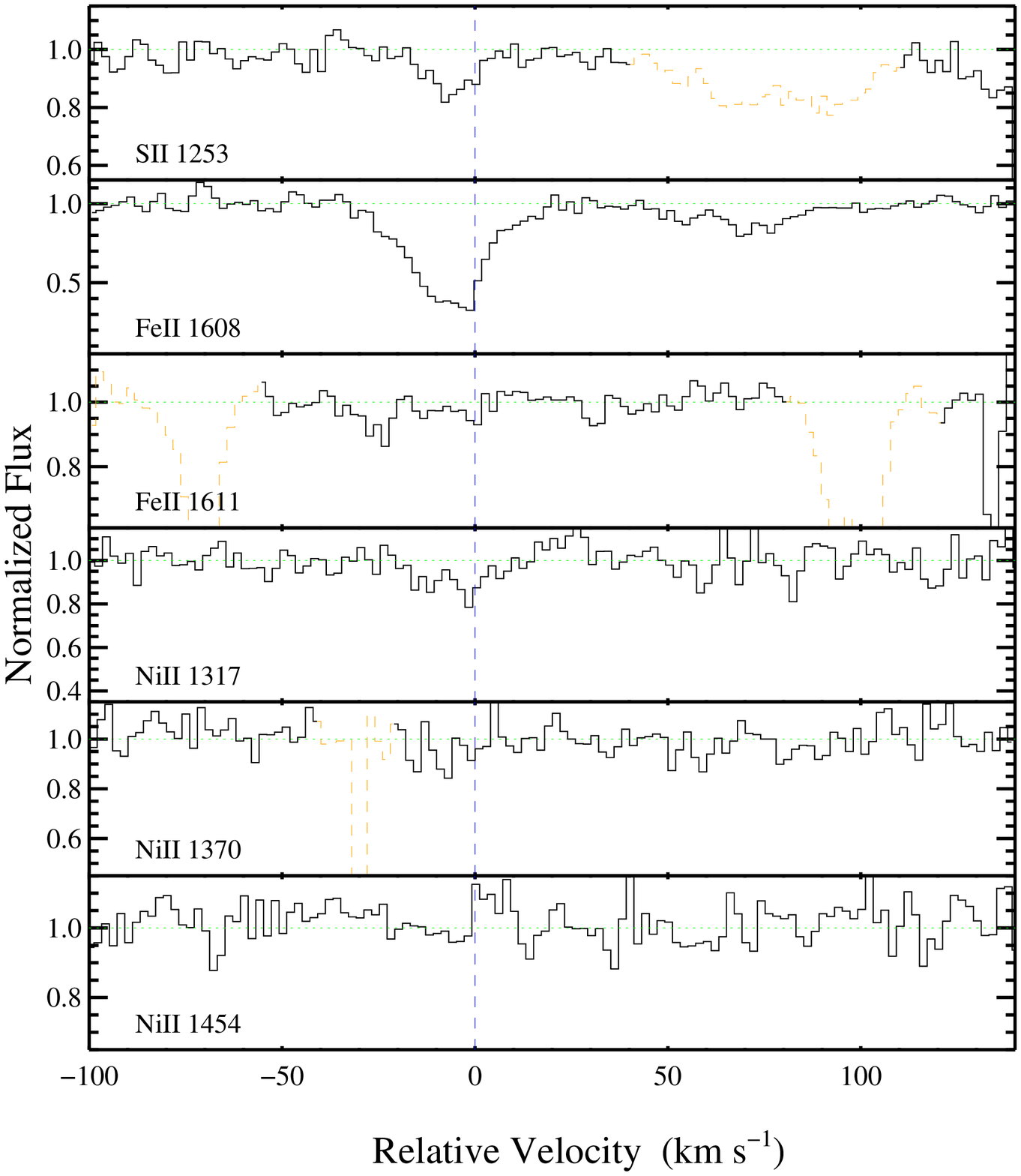}}
\clearpage

\clearpage
\epsscale{0.85}
{\plotone{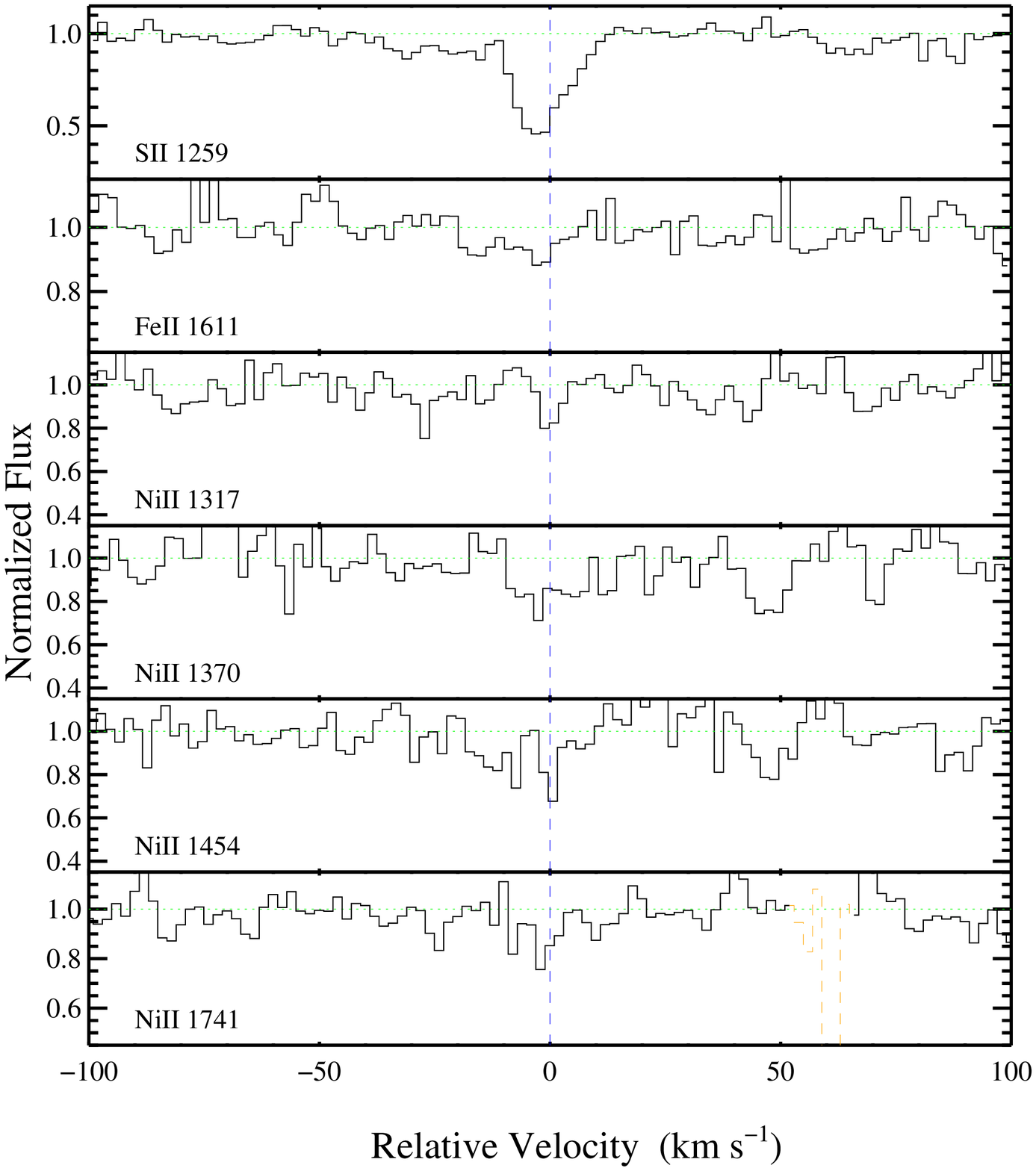}}
\clearpage

\clearpage
\epsscale{0.85}
{\plotone{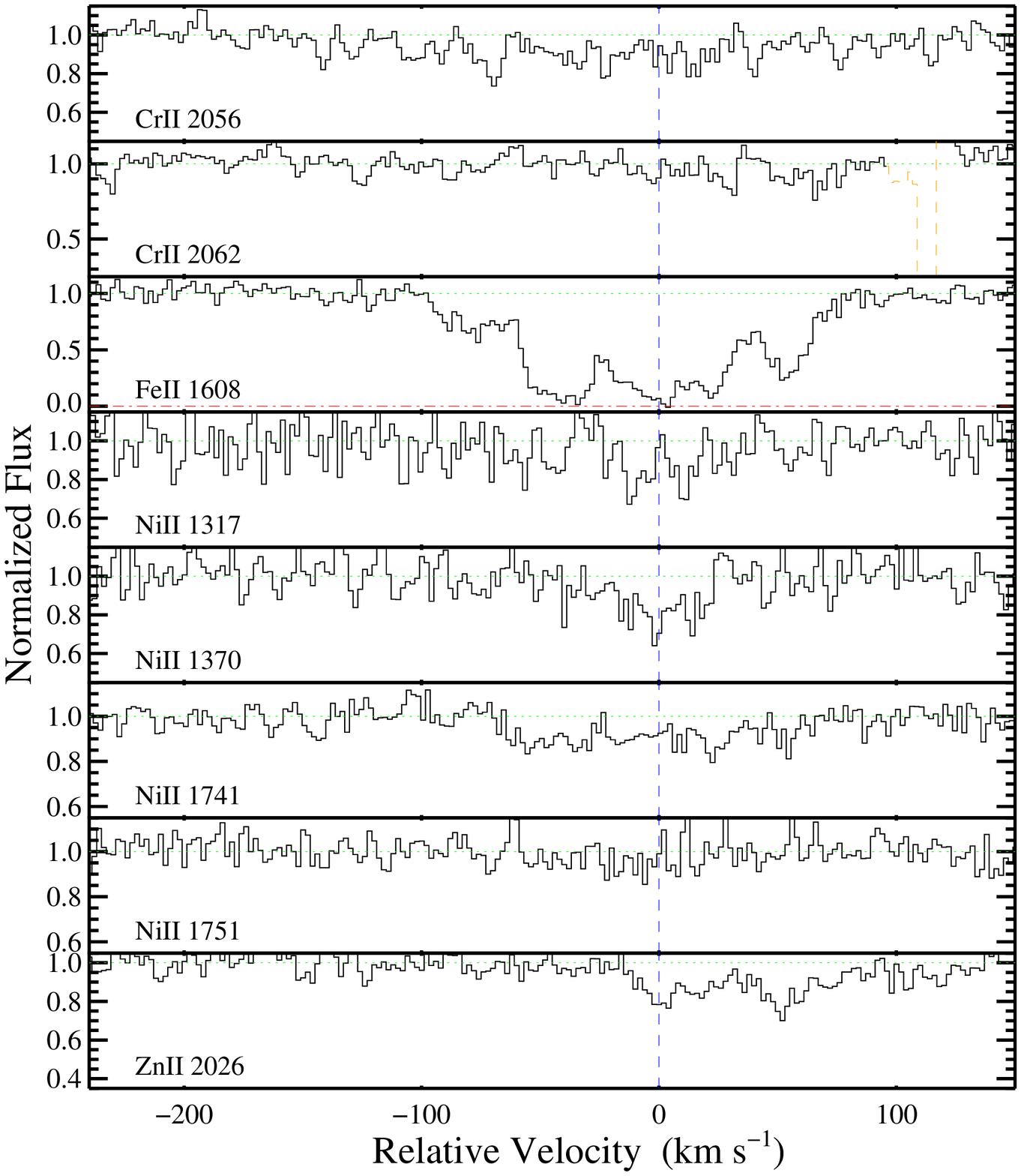}}
\clearpage

\end{document}